\documentclass[english]{article}
\usepackage[T1]{fontenc}
\usepackage[latin9]{inputenc}
\usepackage{verbatim}
\usepackage{amsmath}
\usepackage{amssymb}
\usepackage{graphicx}
\usepackage{esint}

\makeatletter

\providecommand{\tabularnewline}{\\}

\makeatletter
 \@addtoreset{equation}{section}
 
\makeatother
\providecommand{\tabularnewline}{\\}

\allowdisplaybreaks[1]

\newif\ifContLineOne
\newif\ifContLineTwo
\newif\ifContLineThree

\def\conC#1{\vbox{\ialign{##\crcr
  \ifContLineThree\hrulefill\else\vphantom{\hrulefill}\fi\crcr
  \noalign{\kern3.2pt\nointerlineskip}
  \ifContLineTwo\hrulefill\else\vphantom{\hrulefill}\fi\crcr
  \noalign{\kern3.2pt\nointerlineskip}
  \ifContLineOne\hrulefill\else\vphantom{\hrulefill}\fi\crcr
  \noalign{\nointerlineskip}
  $\hfil\textstyle{\vbox to 14pt{}#1}\hfil$\crcr}}}

\def\DrawLeg#1#2{
  \kern-.2pt              
  \dimen2 =#1             
  \advance\dimen2 by 2pt  
  \dimen3 = 10.6pt        
  \dimen4 =3.6pt          
  \advance\dimen3 by -\dimen2 
  \multiply\dimen4 by #2
  \advance\dimen3 by \dimen4
  \raise\dimen2 \hbox{\vrule height\dimen3 width .4pt} 
  \kern-.2pt}             

\def\begC#1#2{\setbox0 =\hbox{$\textstyle{#2}$}
  \dimen0=.5\wd0 \dimen1=\ht0
  \conC{\hskip\dimen0}
  \count255=#1
  \ifnum\count255 =1 \ContLineOnetrue\else
  \ifnum\count255 =2 \ContLineTwotrue\else
  \ifnum\count255 =3 \ContLineThreetrue\fi\fi\fi
  \DrawLeg{\dimen1}{\count255}
  \conC{\hskip\dimen0}
  \kern-\dimen0\kern-\dimen0 \box0}

\def\endC#1#2{\setbox0 =\hbox{$\textstyle{#2}$}
  \dimen0=.5\wd0 \dimen1=\ht0
  \conC{\hskip\dimen0}
  \count255=#1
  \ifnum\count255 =1 \ContLineOnefalse\else
  \ifnum\count255 =2 \ContLineTwofalse\else
  \ifnum\count255 =3 \ContLineThreefalse\fi\fi\fi
  \DrawLeg{\dimen1}{\count255}
  \conC{\hskip\dimen0}
  \kern-\dimen0\kern-\dimen0 \box0}

\makeatother

\usepackage{babel}
\begin{document}
\begin{titlepage}

\global\long\def\thefootnote{\fnsymbol{footnote}}

\begin{flushright}
\begin{tabular}{l}
UTHEP-686 \tabularnewline
\end{tabular}
\par\end{flushright}

\bigskip{}

\begin{center}
\textbf{\Large{}{}{}{}{}Light-cone gauge superstring field theory
in linear dilaton background}{\Large{}{}{}{} } 
\par\end{center}

\bigskip{}

\begin{center}
{\large{}{}{}{}{}{}{}{}Nobuyuki Ishibashi}%
\footnote{e-mail: ishibash@het.ph.tsukuba.ac.jp%
} {\large{}{}{}{}{}{}{}{}}
\par\end{center}{\large \par}

\begin{center}
Center for Integrated Research in Fundamental Science and Engineering
(CiRfSE),\\
 Faculty of Pure and Applied Sciences, University of Tsukuba\\
 Tsukuba, Ibaraki 305-8571, JAPAN
\par\end{center}

\bigskip{}

\bigskip{}

\bigskip{}

\begin{abstract}
The Feynman amplitudes of light-cone gauge superstring field theory
suffer from various divergences. In order to regularize them, we study
the theory in linear dilaton background $\Phi =-iQX^1$ with the number of spacetime
dimensions fixed. We show that the theory with the Feynman $i\varepsilon\,(\varepsilon>0)$
and $Q^{2}>10$ yields finite
results.  
\end{abstract}
\global\long\def\thefootnote{\arabic{footnote}}

\end{titlepage}

\section{Introduction}

Light-cone gauge superstring field theory \cite{Mandelstam:1974hk,Mandelstam:1985wh,Sin:1988yf,Green1983b,Green1983,Gross1987}
was proposed to give a nonperturbative definition of closed superstring
theory with only three-string interaction terms. However it is known
that the Feynman amplitudes of the theory are plagued with various
divergences. Even the tree amplitudes are ill defined because of the
so-called contact term divergences \cite{Greensite:1986gv,Greensite:1987sm,Greensite:1987hm,Green:1987qu}
caused by the insertions of world sheet supercurrents at the interaction
points. 

In our previous works \cite{Baba:2009kr,Baba:2009ns,Baba:2009fi,Baba:2009zm,Ishibashi:2011fy,Ishibashi:2013nma,Ishibashi2016,Ishibashi2011a,Murakami2011},
we have proposed dimensional regularization to deal with the contact
term divergences of the light-cone gauge superstring field theory
in the RNS formalism. Since the light-cone gauge theory is a completely
gauge fixed theory, it is possible to formulate it in $d$ dimensional
Minkowski space with $d\ne10$. Although Lorentz invariance is broken,
the theory corresponds to a conformal gauge world sheet theory with
nonstandard longitudinal part. The world sheet theory for the longitudinal
variables turns out to be a superconformal field theory with the right
central charge so that we can construct nilpotent BRST charge. The
contributions from the longitudinal part of the world sheet theory,
or equivalently the anomaly factors which appear in the light-cone
gauge Feynman amplitudes, have the effect of taming the contact term
divergences and the tree amplitudes become finite when $-d$ is large
enough. It is possible to define the amplitudes as analytic functions
of $d$ and take the limit $d\to10$ to get the amplitudes for critical
strings. The results coincide with those of the first quantized formalism. 

We expect that the dimensional regularization or its variant also
works as a regularization of the multiloop amplitudes. In order to
generalize our results to the multiloop case, there are several things
to be done. We need to study how divergences of multiloop amplitudes
arise in light-cone gauge perturbation theory and check if they are
regularized by considering the theory in noncritical dimension. Another
problem to be considered is how to deal with the spacetime fermions.
Naive dimensional regularization causes problems about spacetime fermions
because the number of gamma matrices is modified in the regularization. 

In this paper, we propose superstring field theory in linear dilaton
background to regularize the divergences of the multiloop amplitudes.
With the linear dilaton background keeping the number of transverse
coordinates to be eight, we can tame the divergences without having
problems about fermions. A Feynman amplitude is given as an integral
over moduli parameters and the integrand is written in terms of quantities
defined on the light-cone diagram. The divergences originate from
degenerations of the world sheet and collisions of interaction points.
We prove that with the Feynman $i\varepsilon\,(\varepsilon>0)$ and
the background charge $Q$ satisfying $Q^{2}>10$, the amplitudes
become finite. It should be possible to define the amplitudes as analytic
functions of $Q$ for $Q^{2}>10$ and take the limit $Q\to0$ to obtain
those in the critical dimension. What happens in the limit is the
subject of another paper \cite{Ishibashi2016a}. 

The organization of this paper is as follows. In section \ref{sec:Superstring-field-theory},
we construct the superstring theory in linear dilaton background and
present the perturbative amplitudes obtained from the theory. In section
\ref{sec:Light-cone-diagrams-in}, we study the light-cone diagrams
which contain divergences. Divergences originate from degeneration
of the world sheet and collisions of interaction points. We study
how light-cone diagrams corresponding to degenerate Riemann surfaces
look like. In section \ref{sec:Degenerate-Riemann-surfaces}, we examine
the singular contributions to the amplitudes from the light-cone diagrams
in which degenerations and collisions of interaction points occur.
In section \ref{sec:Regularization-of-the}, we show that the amplitudes
are finite for the theory with the Feynman $i\varepsilon\,(\varepsilon>0)$
and the background charge $Q$ satisfying $Q^{2}>10$. Section \ref{sec:Discussions}
is devoted to discussions.

\section{Superstring field theory in linear dilaton background\label{sec:Superstring-field-theory}}

\subsection{Light-cone gauge superstring field theory}

In light-cone gauge closed superstring field theory, the string field
\[
\left|\Phi\left(t,\alpha\right)\right\rangle 
\]
 is taken to be an element of the Hilbert space of the transverse
variables on the world sheet and a function of 
\begin{eqnarray*}
t & = & x^{+}\,,\\
\alpha & = & 2p^{+}\,.
\end{eqnarray*}
In this paper, we consider the superstring theory in the RNS formalism.
$\left|\Phi(t,\alpha)\right\rangle $ should be GSO even and satisfy
the level matching condition
\begin{equation}
(L_{0}-\bar{L}_{0})\left|\Phi\left(t,\alpha\right)\right\rangle =0\,,\label{eq:levelmatching}
\end{equation}
where $L_{0},\bar{L}_{0}$ are the Virasoro generators of the world
sheet theory.

In type II superstring theory, the Hilbert space consists of (NS,NS),
(NS,R), (R,NS) and (R,R) sectors and the string fields in the (NS,NS)
and (R,R) sectors are bosonic and those in the (NS,R), (R,NS) sectors
are fermionic. In the heterotic case, there are NS and R sectors of
the right-moving modes, which correspond to bosonic and fermionic
fields respectively. In the following, we will consider the case of
type II theory based on a world sheet theory for the transverse variables
with central charge 
\[
c=\frac{3}{2}(d-2)\,.
\]
The heterotic case can be dealt with in a similar way. 

The action of the string field theory is given by \cite{Baba:2009kr,Ishibashi:2010nq}
\begin{eqnarray}
S & = & \int dt\left[\frac{1}{2}\sum_{\mathrm{B}}\int_{-\infty}^{\infty}\frac{\alpha d\alpha}{4\pi}\left\langle \Phi_{\mathrm{B}}\left(-\alpha\right)\right|(i\partial_{t}-\frac{L_{0}+\bar{L}_{0}-\frac{d-2}{8}-i\varepsilon}{\alpha})\left|\Phi_{\mathrm{B}}\left(\alpha\right)\right\rangle \right.\nonumber \\
 &  & \hphantom{\int dt\quad}+\frac{1}{2}\sum_{\mathrm{F}}\int_{-\infty}^{\infty}\frac{d\alpha}{4\pi}\left\langle \Phi_{\mathrm{F}}\left(-\alpha\right)\right|(i\partial_{t}-\frac{L_{0}+\bar{L}_{0}-\frac{d-2}{8}-i\varepsilon}{\alpha})\left|\Phi_{\mathrm{F}}\left(\alpha\right)\right\rangle \nonumber \\
 &  & \hphantom{\int dt\quad}-\frac{g_{s}}{6}\sum_{\mathrm{B}_{1},\mathrm{B}_{2},\mathrm{B}_{3}}\int\prod_{r=1}^{3}\left(\frac{\alpha_{r}d\alpha_{r}}{4\pi}\right)\delta\left(\sum_{r=1}^{3}\alpha_{r}\right)\left\langle V_{3}\left|\Phi_{\mathrm{B}_{1}}(\alpha_{1})\right.\right\rangle \left|\Phi_{\mathrm{B}_{2}}(\alpha_{2})\right\rangle \left|\Phi_{\mathrm{B}_{3}}(\alpha_{3})\right\rangle \nonumber \\
 &  & \hphantom{\int dt\quad}\left.-\frac{g_{s}}{2}\sum_{\mathrm{B}_{1},\mathrm{F}_{2},\mathrm{F}_{3}}\int\prod_{r=1}^{3}\left(\frac{\alpha_{r}d\alpha_{r}}{4\pi}\right)\delta\left(\sum_{r=1}^{3}\alpha_{r}\right)\left\langle V_{3}\left|\Phi_{\mathrm{B}_{1}}(\alpha_{1})\right.\right\rangle \alpha_{2}^{-\frac{1}{2}}\left|\Phi_{\mathrm{F}_{2}}(\alpha_{2})\right\rangle \alpha_{3}^{-\frac{1}{2}}\left|\Phi_{\mathrm{F}_{3}}(\alpha_{3})\right\rangle \right]\,.\nonumber \\
 &  & \ \label{eq:superHamiltonian}
\end{eqnarray}
 The first and the second terms are the kinetic terms with the Feynman
$i\varepsilon$ and $\left\langle \Phi(-\alpha)\right|$ denotes the
BPZ conjugate of $\left|\Phi(-\alpha)\right\rangle $. The third and
the fourth terms are the three string vertices and $g_{s}$ is the
string coupling constant. $\sum_{\mathrm{B}}$ and $\sum_{\mathrm{F}}$
denote the sums over bosonic and fermionic string fields respectively. 

\begin{figure}
\begin{centering}
\includegraphics[scale=0.8]{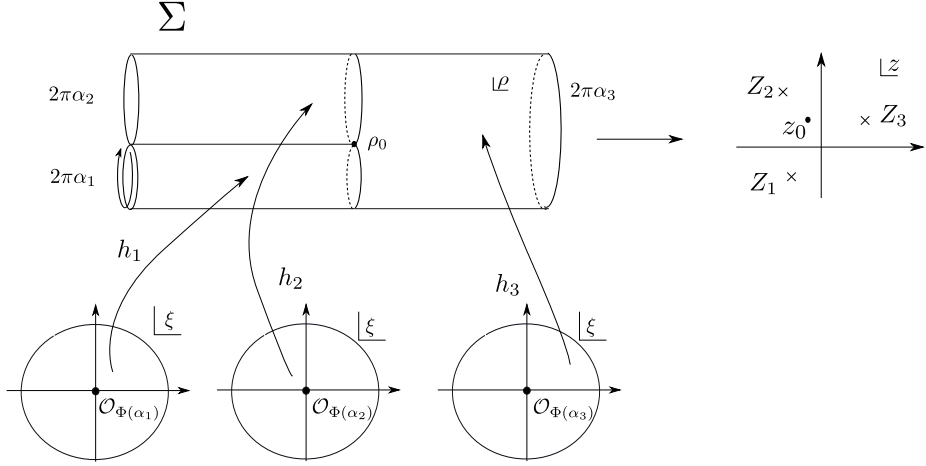}
\par\end{centering}

\protect\caption{The three string vertex for superstrings. Here we consider the case
$\alpha_{1},\alpha_{2}>0,\,\alpha_{3}<0$.\label{fig:The-three-string} }
\end{figure}

By the state-operator correspondence of the world sheet conformal
field theory, there exists a local operator $\mathcal{O}_{\Phi}(\xi,\bar{\xi})$
corresponding to any state $\left|\Phi\right\rangle $. We define
$\left\langle \left.V_{3}\right|\Phi(\alpha_{1})\right\rangle \left|\Phi(\alpha_{2})\right\rangle \left|\Phi(\alpha_{3})\right\rangle $
with $\sum_{r=1}^{3}\alpha_{r}=0$ to be
\begin{eqnarray}
 &  & \left\langle \left.V_{3}\right|\Phi(\alpha_{1})\right\rangle \left|\Phi(\alpha_{2})\right\rangle \left|\Phi(\alpha_{3})\right\rangle \nonumber \\
 &  & \quad=\left\langle \lim_{\rho\to\rho_{0}}\left|\rho-\rho_{0}\right|^{\frac{3}{2}}T_{F}^{LC}\left(\rho\right)\bar{T}_{F}^{LC}\left(\bar{\rho}\right)h_{1}\circ\mathcal{O}_{\Phi\left(\alpha_{1}\right)}(0,0)h_{2}\circ\mathcal{O}_{\Phi\left(\alpha_{2}\right)}(0,0)h_{3}\circ\mathcal{O}_{\Phi\left(\alpha_{3}\right)}(0,0)\right\rangle _{\Sigma}\,,\label{eq:bosonicthree}
\end{eqnarray}
in terms of a correlation function on $\Sigma$ which is the world
sheet describing the three string interaction depicted in Fig. \ref{fig:The-three-string}.
On each cylinder corresponding to an external line, one can introduce
a complex coordinate 
\[
\rho=\tau+i\sigma\,,
\]
 whose real part $\tau$ coincides with the Wick rotated light-cone
time $it$ and imaginary part $\sigma\sim\sigma+2\pi\alpha_{r}$ parametrizes
the closed string at each time. The $\rho$'s on the cylinders are
smoothly connected except at the interaction point $\rho_{0}$ and
we get a complex coordinate $\rho$ on $\Sigma$. The correlation
function $\left\langle \cdot\right\rangle _{\Sigma}$ is defined with
the metric 
\[
ds^{2}=d\rho d\bar{\rho}\,,
\]
on the world sheet. $h_{r}(\xi)$ gives a map from a unit disk $\left|\xi\right|<1$
to the cylinder corresponding to the $r$-th external line so that
\[
\xi=e^{\frac{1}{\alpha_{r}}(h_{r}(\xi)-\rho_{0})}\,.
\]
 $T_{F}^{LC},\bar{T}_{F}^{LC}$ are the supercurrents of the transverse
world sheet theory. 

It is convenient to express the right hand side of (\ref{eq:bosonicthree})
in terms of a correlation function on the sphere as 
\begin{eqnarray}
 &  & \left\langle \left.V_{3}\right|\Phi(\alpha_{1})\right\rangle \left|\Phi(\alpha_{2})\right\rangle \left|\Phi(\alpha_{3})\right\rangle \nonumber \\
 &  & \quad=e^{-\frac{d-2}{16}\Gamma}\left\langle \left|\partial^{2}\rho\left(z_{0}\right)\right|^{-\frac{3}{2}}T_{F}^{LC}\left(z_{0}\right)\bar{T}_{F}^{LC}\left(\bar{z}_{0}\right)\right.\nonumber \\
 &  & \hphantom{\quad=e^{-\frac{d-2}{16}\Gamma}T_{F}^{LC}\left(z_{0}\right)}\times\left.\rho^{-1}h_{1}\circ\mathcal{O}_{\Phi_{1}\left(\alpha_{1}\right)}(0,0)\rho^{-1}h_{2}\circ\mathcal{O}_{\Phi_{2}\left(\alpha_{2}\right)}(0,0)\rho^{-1}h_{3}\circ\mathcal{O}_{\Phi_{3}\left(\alpha_{3}\right)}(0,0)\right\rangle _{\mathbb{C}\cup\infty}\,.\label{eq:superthreestring}
\end{eqnarray}
Here $\rho(z)$ is given by 
\[
\rho(z)=\sum_{r=1}^{3}\alpha_{r}\ln(z-Z_{r})\,,
\]
which maps the complex plane to $\Sigma$. $z_{0}$ denotes the $z$-coordinate
of the interaction point, which satisfies
\[
\partial\rho(z_{0})=0\,.
\]
The correlation function $\left\langle \cdot\right\rangle _{\mathbb{C}\cup\infty}$
is defined with the metric
\[
ds^{2}=dzd\bar{z}\,,
\]
on the world sheet. The salient feature of light-cone gauge string
field theory is that the central charge of the world sheet theory
is nonvanishing even in the critical case. $e^{-\frac{d-2}{16}\Gamma}$
is the anomaly factor associated to the conformal map $z\to\rho(z)$
and its explicit form is given as 
\[
e^{-\frac{d-2}{16}\Gamma}=\left(\frac{\exp\left(-2\sum_{r}\frac{\hat{\tau}_{0}}{\alpha_{r}}\right)}{\alpha_{1}\alpha_{2}\alpha_{3}}\right)^{\frac{d-2}{16}}\,,
\]
where 
\[
\hat{\tau}_{0}\equiv\sum_{r=1}^{3}\alpha_{r}\ln\left|\alpha_{r}\right|\,.
\]

\subsection{Linear dilaton background\label{sec:Worldsheet-theory}}

The Feynman amplitudes of light-cone gauge superstring field theory
suffer from the contact term divergences. As we have pointed out in
\cite{Baba:2009kr}, these divergences are regularized by formulating
the theory in $d\ne10$ dimensions. However, if one simply considers
superstring field theory in noncritical dimensions, fermionic string
fields cannot satisfy (\ref{eq:levelmatching}) \cite{Ishibashi:2011fy}.
In type II superstring theory, naive dimensional continuation implies
that the level-matching condition for the (NS,R) sector becomes
\[
\mathcal{N}+\frac{d-2}{16}=\bar{\mathcal{N}}\,,
\]
where $\mathcal{N}$ and $\bar{\mathcal{N}}$ denote the left and
right mode numbers of the light-cone gauge string state. For generic
$d$, there exists no state satisfying it. The same argument applies
to (R,NS) sector. We have the same problem in the R sector of the
heterotic string theory. Therefore we cannot use the naive dimensional
regularization to regularize superstring amplitudes, although it may
be used to deal with the type 0 theories. Another drawback of the
naive dimensional regularization is that there is difficulty in dealing
with odd spin structure %
\footnote{The world sheet theory proposed in \cite{Ishibashi:2011fy} has this
problem. %
}, as anticipated from the problems with $\gamma^{5}$ in the dimensional
regularization in field theory. 

It is possible to regularize the divergences by formulating the theory
based on world sheet theory with a large negative central charge,
instead of changing the number of spacetime dimensions. Therefore,
in order to have a theory without the aforementioned problems, we
need a world sheet theory in which we can change the central charge
keeping the number of the world sheet fermions fixed. A convenient
way to obtain such a theory is to take the dilaton background to be
$\Phi =-iQX^1$, proportional to one of the transverse target space coordinates 
$X^{1}$. Then the world sheet action of $X^{1}$ and its fermionic
partners $\psi^{1},\bar{\psi^{1}}$ on a world sheet with metric $ds^{2}=2\hat{g}_{z\bar{z}}dzd\bar{z}$
becomes
\begin{eqnarray}
S\left[X^{1},\psi^{1},\bar{\psi}^{1};\hat{g}_{z\bar{z}}\right] & = & \frac{1}{16\pi}\int dz\wedge d\bar{z}i\sqrt{\hat{g}}\left(\hat{g}^{ab}\partial_{a}X^{1}\partial_{b}X^{1}-2iQ\hat{R}X^{1}\right)\nonumber \\
 &  & \qquad+\frac{1}{4\pi}\int dz\wedge d\bar{z}i\left(\psi^{1}\bar{\partial}\psi^{1}+\bar{\psi}^{1}\partial\bar{\psi}^{1}\right)\,,\label{eq:linaction}
\end{eqnarray}
and the energy-momentum tensor and the supercurrent are given as
\begin{eqnarray*}
T^{X^{1}}(z) & = & -\frac{1}{2}(\partial X^{1})^{2}-iQ(\partial-\partial\ln\hat{g}_{z\bar{z}})\partial X^{1}-\frac{1}{2}\psi^{1}\partial\psi^{1}\,,\\
T_{F}^{X^{1}}(z) & = & -\frac{i}{2}\partial X^{1}\psi^{1}+Q(\partial-\frac{1}{2}\partial\ln\hat{g}_{z\bar{z}})\psi^{1}\,.
\end{eqnarray*}
In this paper, we take $Q$ to be a real constant. 

In order to construct string field theory and calculate amplitudes
we need the correlation functions of the linear dilaton conformal
field theory. Since the fermionic part is just a free theory we concentrate
on the bosonic part. Let us consider the correlation function of operators
$e^{ip_{r}X^{1}}\,(r=1,\cdots,N)$ on a Riemann surface of genus $g$
with metric $ds^{2}=2g_{z\bar{z}}dzd\bar{z}$, which is given as
\begin{equation}
\int\left[dX^{1}\right]_{g_{z\bar{z}}}e^{-S\left[X^{1};g_{z\bar{z}}\right]}\prod_{r=1}^{N}e^{ip_{r}X^{1}}(Z_{r},\bar{Z}_{r})\,,\label{eq:linearcorrelation}
\end{equation}
where
\[
S\left[X^{1};g_{z\bar{z}}\right]=\frac{1}{16\pi}\int dz\wedge d\bar{z}i\sqrt{g}\left(g^{ab}\partial_{a}X^{1}\partial_{b}X^{1}-2iQRX^{1}\right)\,.
\]
We would like to express (\ref{eq:linearcorrelation}) in terms of
the correlation function on the world sheet with a fiducial metric
$ds^{2}=2\hat{g}_{z\bar{z}}dzd\bar{z}$. It is straightforward to
show 
\begin{eqnarray*}
 &  & \int\left[dX^{1}\right]_{g_{z\bar{z}}}e^{-S\left[X^{1};g_{z\bar{z}}\right]}\prod_{r=1}^{N}e^{ip_{r}X^{1}}(Z_{r},\bar{Z}_{r})\\
 &  & \quad=e^{-\frac{1-12Q^{2}}{24}\Gamma\left[\sigma;\hat{g}_{z\bar{z}}\right]}\int\left[d\hat{X}{}^{1}\right]_{\hat{g}_{z\bar{z}}}e^{-S\left[\hat{X}^{1};\hat{g}_{z\bar{z}}\right]}\prod_{r=1}^{N}\left[e^{ip_{r}\hat{X}^{1}}\left(\frac{g_{z\bar{z}}}{\hat{g}_{z\bar{z}}}\right)^{-Qp_{r}}\right](Z_{r},\bar{Z}_{r})\,,
\end{eqnarray*}
where 
\begin{eqnarray}
\sigma & \equiv & \ln g_{z\bar{z}}-\ln\hat{g}_{z\bar{z}}\,,\nonumber \\
\Gamma\left[\sigma;\hat{g}_{z\bar{z}}\right] & = & -\frac{1}{4\pi}\int dz\wedge d\bar{z}i\sqrt{\hat{g}}\left(\hat{g}^{ab}\partial_{a}\sigma\partial_{b}\sigma+2\hat{R}\sigma\right)\,,\label{eq:Gamma}\\
\hat{X}^{1} & \equiv & X^{1}-iQ\sigma\,.\nonumber 
\end{eqnarray}
The anomaly factor $e^{-\frac{1-12Q^{2}}{24}\Gamma\left[\sigma;\hat{g}_{z\bar{z}}\right]}$
is exactly what we expect for a theory with the central charge
\[
c=1-12Q^{2}
\]
 of the linear dilaton conformal field theory. The correlation functions
with the fiducial metric $ds^{2}=2\hat{g}_{z\bar{z}}dzd\bar{z}$ can
be calculated by introducing the Arakelov Green's function $G^{\mathrm{A}}(z,w)$
\cite{arakelov} which satisfies
\begin{eqnarray*}
 &  & \partial_{z}\partial_{\bar{z}}G^{\mathrm{A}}(z,w)=-\pi\delta^{2}(z-w)-\frac{\hat{g}_{z\bar{z}}\hat{R}}{4(g-1)}\,,\\
 &  & \int dz\wedge d\bar{z}i\sqrt{\hat{g}}\hat{R}G^{\mathrm{A}}(z,w)=0\,,
\end{eqnarray*}
and the result is 
\begin{eqnarray*}
 &  & \int\left[d\hat{X}^{1}\right]_{\hat{g}_{z\bar{z}}}e^{-S\left[\hat{X}^{1};\hat{g}_{z\bar{z}}\right]}\prod e^{ip_{r}^{1}\hat{X}}(Z_{r},\bar{Z}_{r})\\
 &  & \quad=2\pi\delta\left(\sum p_{r}+2Q(1-g)\right)Z^{X}\left[\hat{g}_{z\bar{z}}\right]\\
 &  & \hphantom{\quad=\quad}\times\prod_{r>s}e^{-p_{r}p_{s}G^{\mathrm{A}}(Z_{r},Z_{s})}\prod_{r}e^{-\frac{1}{2}p_{r}^{2}\lim_{z\to Z_{r}}(G^{\mathrm{A}}(z,Z_{r})+\ln\left|z-Z_{r}\right|^{2})}\,.
\end{eqnarray*}
Here $Z^{X}\left[\hat{g}_{z\bar{z}}\right]$ denotes the partition
function of a free scalar on the world sheet with metric $ds^{2}=2\hat{g}_{z\bar{z}}dzd\bar{z}$.
Taking the fiducial metric to be the Arakelov metric $g_{z\bar{z}}^{\mathrm{A}}$
\cite{arakelov}, for which the Arakelov Green's function $G^{\mathrm{A}}(z,w)$
satisfies 
\[
\lim_{w\to z}\left(G^{\mathrm{A}}(w,z)+\ln\left|z-w\right|^{2}\right)=-\ln\left(2g_{z\bar{z}}^{\mathrm{A}}\right)\,,
\]
the correlation function becomes
\[
2\pi\delta\left(\sum p_{r}+2Q(1-g)\right)Z^{X}\left[g_{z\bar{z}}^{\mathrm{A}}\right]\prod_{r>s}e^{-p_{r}p_{s}G^{\mathrm{A}}(Z_{r},Z_{s})}\prod_{r}\left(2g_{Z_r\bar{Z}_r}^{\mathrm{A}}\right)^{\frac{1}{2}p_{r}^{2}}\,.
\]
From these calculations, we can see that it is convenient to define
\[
\tilde{X}^{1}\equiv X^{1}-iQ\ln(2g_{z\bar{z}})\,,
\]
so that the correlation function of $e^{ip_{r}\tilde{X}^{1}}(Z_{r},\bar{Z}_{r})\,(r=1,\cdots,N)$
is expressed as 
\begin{eqnarray}
 &  & \int\left[dX^{1}\right]_{g_{z\bar{z}}}e^{-S\left[X^{1};g_{z\bar{z}}\right]}\prod_{r=1}^{N}e^{ip_{r}\tilde{X}^{1}}(Z_{r},\bar{Z}_{r})\nonumber \\
 &  & \quad=2\pi\delta\left(\sum p_{r}+2Q(1-g)\right)e^{-\frac{1-12Q^{2}}{24}\Gamma\left[\sigma;g_{z\bar{z}}^{\mathrm{A}}\right]}Z^{X}\left[g_{z\bar{z}}^{\mathrm{A}}\right]\prod_{r>s}e^{-p_{r}p_{s}G^{\mathrm{A}}(Z_{r},Z_{s})}\prod_{r}\left(2g_{Z_r\bar{Z}_r}^{\mathrm{A}}\right)^{\frac{1}{2}p_{r}^{2}+Qp_{r}}\,.\nonumber \\
 &  & \ \label{eq:linearcorr}
\end{eqnarray}
On the sphere, this becomes
\begin{eqnarray}
 &  & \int\left[dX^{1}\right]_{g_{z\bar{z}}}e^{-S\left[X^{1};g_{z\bar{z}}\right]}\prod_{r=1}^{N}e^{ip_{r}\tilde{X}^{1}}(Z_{r},\bar{Z}_{r})\nonumber \\
 &  & \quad=2\pi\delta\left(\sum p_{r}+2Q\right)e^{-\frac{1-12Q^{2}}{24}\Gamma\left[\sigma;g_{z\bar{z}}^{\mathrm{A}}\right]}\prod_{r>s}\left|Z_{r}-Z_{s}\right|^{2p_{r}p_{s}}\,.\label{eq:lincorrtree}
\end{eqnarray}

$e^{ip\tilde{X}^{1}}$ thus defined turns out to be a primary field
with conformal dimension 
\begin{equation}
\frac{1}{2}p^{2}+Qp=\frac{1}{2}(p+Q)^{2}-\frac{Q^{2}}{2}\,.\label{eq:lineardilatondim}
\end{equation}
Notice that $\tilde{X}^{1}$ satisfies
\[
\partial\bar{\partial}\tilde{X}^{1}=0\,,
\]
if there are no source terms and $i\partial\tilde{X}^{1}(z),i\bar{\partial}\tilde{X}^{1}(\bar{z})$
can be expanded as
\begin{eqnarray*}
i\partial\tilde{X}^{1}(z) & = & \sum_{n}\alpha_{n}^{1}z^{-n-1}\,,\\
i\bar{\partial}\tilde{X}^{1}(\bar{z}) & = & \sum_{n}\bar{\alpha}_{n}^{1}\bar{z}^{-n-1}\,,
\end{eqnarray*}
where $\alpha_{n}^{1}$ and $\bar{\alpha}_{n}^{1}$ satisfy the canonical
commutation relations. The states in the CFT can be expressed as linear
combinations of the Fock space states 
\begin{equation}
\alpha_{-n_{1}}^{1}\cdots\alpha_{-n_{k}}^{1}\bar{\alpha}_{-\bar{n}_{1}}^{1}\cdots\bar{\alpha}_{-\bar{n}_{l}}^{1}\left|p\right\rangle \,.\label{eq:linearfock}
\end{equation}
It is straightforward to show
\begin{eqnarray*}
\langle p_{1}|p_{2}\rangle & = & 2\pi\delta(p_{1}+p_{2}+2Q)\,,\\
\left(\alpha_{n}^{1}\right)^{\ast} & = & (-1)^{n+1}(\alpha_{-n}^{1}+2Q\delta_{n,0})\,,\\
\left(\bar{\alpha}_{n}^{1}\right)^{\ast} & = & (-1)^{n+1}(\bar{\alpha}_{-n}^{1}+2Q\delta_{n,0})\,,
\end{eqnarray*}
where $\left\langle p\right|,\left(\alpha_{n}^{1}\right)^{\ast},\left(\bar{\alpha}_{n}^{1}\right)^{\ast}$
are the BPZ conjugates of $\left|p\right\rangle ,\alpha_{n}^{1},\bar{\alpha}_{n}^{1}$
respectively. %

\subsection{Light-cone gauge superstring field theory in linear dilaton background}

\begin{figure}
\begin{centering}
\includegraphics[scale=0.5]{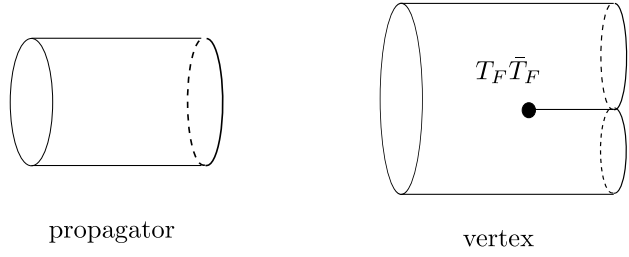}
\par\end{centering}

\protect\caption{Propagator and vertex.\label{fig:Propagator-and-vertex.}}

\end{figure}

\begin{figure}
\begin{centering}
\includegraphics[scale=1.4]{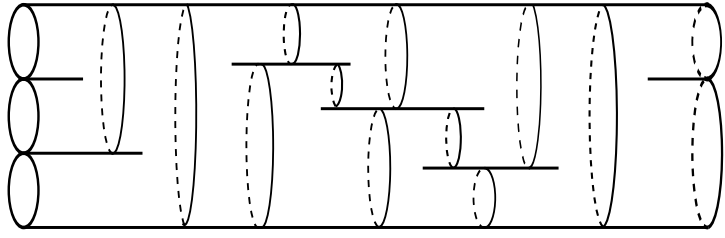}
\par\end{centering}

\protect\caption{A string diagram with $3$ incoming, $2$ outgoing strings and $3$
loops. \label{fig:A-string-diagram}}
\end{figure}

Now let us construct the light-cone gauge superstring field theory
based on the world sheet theory with the variables 
\[
X^{i},\psi^{i},\bar{\psi}^{i}\ (i=1,\cdots,8)\,,
\]
where the action for $X^{1},\psi^{1},\bar{\psi}^{1}$ is taken to
be (\ref{eq:linaction}) and that for other variables is the free
one. The world sheet theory of the transverse variables turns out
to be a superconformal field theory with central charge 
\[
c=12-12Q^{2}\,.
\]
Therefore we can make $-c$ arbitrarily large keeping the number of
the transverse fermionic variables fixed. From the correlation functions
(\ref{eq:lincorrtree}) of the conformal field theory, it is straightforward
to construct the light-cone gauge string field action (\ref{eq:superHamiltonian})
with 
\[
d-2=8-8Q^{2}\,.
\]
The Feynman amplitudes are calculated by the old-fashioned perturbation
theory starting from the action (\ref{eq:superHamiltonian}). Each
term in the expansion corresponds to a light-cone gauge Feynman diagram
for strings which is constructed from the propagator and the vertex
presented in Fig. \ref{fig:Propagator-and-vertex.}. A typical diagram
is depicted in Fig. \ref{fig:A-string-diagram}.

A Wick rotated $g$-loop $N$-string diagram is conformally equivalent
to an $N$ punctured genus $g$ Riemann surface $\Sigma_{N}^{(g)}$.
A light-cone diagram consists of cylinders which correspond to propagators
of the closed string. On each cylinder, one can introduce a complex
coordinate $\rho$ as we did for the three string vertex. The $\rho$'s
on the cylinders are smoothly connected except at the interaction
points and we get a complex coordinate $\rho$ on $\Sigma_{N}^{(g)}$.
$\rho$ is not a good coordinate at the punctures and the interaction
points. 

$\rho$ can be given as a function of a local coordinate $z$ on $\Sigma_{N}^{(g)}$
as \cite{Ishibashi:2013nma} 
\begin{equation}
\rho(z)=\sum_{r=1}^{N}\alpha_{r}\left[\ln E(z,Z_{r})-2\pi i\int_{P_{0}}^{z}\omega\frac{1}{\mathop{\mathrm{Im}}\Omega}\mathop{\mathrm{Im}}\int_{P_{0}}^{Z_{r}}\omega\right]~,\qquad\sum_{r=1}^{N}\alpha_{r}=0~,\label{eq:mandelstammulti}
\end{equation}
up to an additive constant independent of $z$. Here $E(z,w)$ is
the prime form, $\omega$ is the canonical basis of the holomorphic
abelian differentials and $\Omega$ is the period matrix of the surface.%
\footnote{For the mathematical background relevant for string perturbation theory,
we refer the reader to \cite{D'Hoker:1988ta}. %
} The base point $P_{0}$ is arbitrary. There are $2g-2+N$ zeros of
$\partial\rho$ and we denote them by $z_{I}\,(I=1,\cdots,2g-2+N)$.
They correspond to the interaction points of the light-cone diagram.

A $g$-loop $N$-string amplitude is given as an integral over the
moduli space of the string diagram as
\begin{equation}
A_{N}^{(g)}=\left(2\pi\right)^{2}\delta\left(\sum_{r=1}^{N}p_{r}^{+}\right)\delta\left(\sum_{r=1}^{N}p_{r}^{-}\right)(ig_{s})^{2g-2+N}C\sum_{\mathrm{channels}}\int\left[dT\right]\left[\alpha d\theta\right]\left[d\alpha\right]F_{N}^{(g)}\,,\label{eq:ANlin}
\end{equation}
where $\int[dT][\alpha d\theta][d\alpha]$ denotes the integration
over the moduli parameters and $C$ is the combinatorial factor. In
each channel, the integration measure is given as \cite{D'Hoker:1987pr}
\begin{equation}
\int[dT][\alpha d\theta][d\alpha]=\prod_{a=1}^{2g-3+N}\left(-i\int_{0}^{\infty}dT_{a}\right)\prod_{A=1}^{g}\int\frac{d\alpha_{A}}{4\pi}\prod_{\mathcal{I}=1}^{3g-3+N}\left(|\alpha_{\mathcal{I}}|\int_{0}^{2\pi}\frac{d\theta_{\mathcal{I}}}{2\pi}\right).\label{eq:measure}
\end{equation}
Here $T_{a}$'s are heights of the cylinders corresponding to internal
lines,%
\footnote{Heights of the cylinders in a light-cone diagram are constrained so
that only $2g-3+N$ of them can be varied independently.%
} $\alpha_{A}$'s denote the widths of the cylinders corresponding
to the $+$ components of the loop momenta and $\alpha_{\mathcal{I}}$'s
and $\theta_{\mathcal{I}}$'s are the string-lengths and the twist
angles of the internal propagators. Summing over channels, with the
natural range of these coordinates, the moduli space%
\footnote{The amplitude (\ref{eq:ANlin}) for $d=10$ can formally be recast
into an integral over the supermoduli space \cite{Berkovits:1985ji,Berkovits:1987gp,Berkovits:1988ve,Berkovits:1988vf,Berkovits:1988xq,Aoki:1990yn}. %
} of the Riemann surface is covered exactly once \cite{Giddings:1986rf}. 

The integrand $F_{N}^{(g)}$ in (\ref{eq:ANlin}) is given as
\[
F_{N}^{(g)}=\int\left[dX^{i}d\psi^{i}d\bar{\psi}^{i}\right]_{g_{z\bar{z}}}e^{-S\left[X^{i},\psi^{i},\bar{\psi}^{i};g_{z\bar{z}}^{\mathrm{A}}\right]}\prod_{I=1}^{2g-2+N}\left(\left|\partial^{2}\rho\left(z_{I}\right)\right|^{-\frac{3}{2}}T_{F}^{\mathrm{LC}}\left(z_{I}\right)\bar{T}_{F}^{\mathrm{LC}}\left(\bar{z}_{I}\right)\right)\prod_{r=1}^{N}V_{r}^{\mathrm{LC}}(Z_{r},\bar{Z}_{r})\,.
\]
Here $V_{r}^{\mathrm{LC}}(Z_{r},\bar{Z}_{r})$ denotes the vertex
operator \cite{Baba:2009kr} for the $r$-th external line and the
insertions of the world sheet supercharges $T_{F}^{\mathrm{LC}}\left(z_{I}\right),\,\bar{T}_{F}^{\mathrm{LC}}\left(\bar{z}_{I}\right)$
originate from those in the three string vertex (\ref{eq:superthreestring}).
The path integral is defined with the world sheet metric 
\begin{equation}
ds^{2}=2g_{z\bar{z}}dzd\bar{z}\equiv\partial\rho\bar{\partial}\bar{\rho}dzd\bar{z}\,.\label{eq:rhometric}
\end{equation}

Since $g_{z\bar{z}}$ is singular at $z=z_{I},Z_{r}$, we need to
rewrite the path integral in terms of that defined with a metric which
is regular everywhere on the world sheet. Taking the world sheet metric
to be the Arakelov metric, we get
\begin{eqnarray}
F_{N}^{(g)} & = & e^{-\frac{1-Q^{2}}{2}\Gamma\left[\sigma;g_{z\bar{z}}^{\mathrm{A}}\right]}\int\left[dX^{i}d\psi^{i}d\bar{\psi}^{i}\right]_{g_{z\bar{z}}^{\mathrm{A}}}e^{-S\left[X^{i},\psi^{i},\bar{\psi}^{i};g_{z\bar{z}}^{\mathrm{A}}\right]}\nonumber \\
 &  & \hphantom{e^{-\frac{1-Q^{2}}{2}\Gamma\left[\sigma;g_{z\bar{z}}^{\mathrm{A}}\right]}\int\quad}\times\prod_{I=1}^{2g-2+N}\left(\left|\partial^{2}\rho\left(z_{I}\right)\right|^{-\frac{3}{2}}T_{F}^{\mathrm{LC}}\left(z_{I}\right)\bar{T}_{F}^{\mathrm{LC}}\left(\bar{z}_{I}\right)\right)\prod_{r=1}^{N}V_{r}^{\mathrm{LC}}(Z_{r},\bar{Z}_{r})\,.\nonumber \\
 &  & \ \label{eq:FNlin}
\end{eqnarray}

It is possible to calculate the quantities which appear on the right
hand side of (\ref{eq:FNlin}). Substituting (\ref{eq:rhometric})
into (\ref{eq:Gamma}) yields a divergent result for $\Gamma\left[\sigma;g_{z\bar{z}}^{\mathrm{A}}\right]$.
We can obtain $e^{-\Gamma\left[\sigma;g_{z\bar{z}}^{\mathrm{A}}\right]}$
up to a divergent numerical factor by regularizing it as was done
in \cite{Mandelstam:1985ww}. The divergent factor can be absorbed
in a redefinition of $g_{s}$ and the vertex operator. $e^{-\Gamma\left[\sigma;g_{z\bar{z}}^{\mathrm{A}}\right]}$
for higher genus surfaces is calculated in \cite{Ishibashi:2013nma}
to be 
\[
e^{-\Gamma\left[\sigma;g_{z\bar{z}}^{\mathrm{A}}\right]}\propto e^{-W}\prod_{r}e^{-2\mathop{\mathrm{Re}}\bar{N}_{00}^{rr}}\prod_{I}\left|\partial^{2}\rho\left(z_{I}\right)\right|^{-3}\,,
\]
up to a numerical constant which can be fixed by imposing the factorization
condition \cite{Ishibashi:2013nma}. Here 
\begin{eqnarray*}
-W & \equiv & -2\sum_{I<J}G^{\mathrm{A}}\left(z_{I};z_{J}\right)-2\sum_{r<s}G^{\mathrm{A}}\left(Z_{r};Z_{s}\right)+2\sum_{I,r}G^{\mathrm{A}}\left(z_{I};Z_{r}\right)\\
 &  & {}-\sum_{r}\ln\left(2g_{Z_{r}\bar{Z}_{r}}^{\mathrm{A}}\right)+3\sum_{I}\ln\left(2g_{z_{I}\bar{z}_{I}}^{\mathrm{A}}\right)\,.\\
\bar{N}_{00}^{rr} & \equiv & \lim_{z\to Z_{r}}\left[\frac{\rho(z_{I^{(r)}})-\rho(z)}{\alpha_{r}}+\ln(z-Z_{r})\right]\\
 & = & \frac{\rho(z_{I^{(r)}})}{\alpha_{r}}-\sum_{s\neq r}\frac{\alpha_{s}}{\alpha_{r}}\ln E(Z_{r},Z_{s})+\frac{2\pi i}{\alpha_{r}}\int_{P_{0}}^{Z_{r}}\omega\frac{1}{\mathop{\mathrm{Im}}\Omega}\sum_{s=1}^{N}\alpha_{s}\mathop{\mathrm{Im}}\int_{P_{0}}^{Z_{s}}\omega~,
\end{eqnarray*}
and $z_{I^{(r)}}$ denotes the coordinate of the interaction point
at which the $r$-th external line interacts. The correlation functions
of $X^{1}$ which appear in (\ref{eq:FNlin}) can be derived from
(\ref{eq:linearcorr}). $Z^{X}\left[g_{z\bar{z}}^{\mathrm{A}}\right]$
and the correlation functions involving other variables have been
calculated on higher genus Riemann surfaces in \cite{AlvarezGaume:1987vm,Verlinde:1986kw,Dugan:1987qe,Sonoda1987c,Atick1987b}. 

From the explicit form of these quantities, we can see that $F_{N}^{(g)}$
could become singular if and only if either or both of the following
things happen:
\begin{enumerate}
\item Some of the interaction points collide with each other.
\item The Riemann surface corresponding to the world sheet degenerates.%
\footnote{The case in which a puncture and an interaction point collide is included
in this category, because of the identity \cite{Sonoda:1987ra} 
\[
\left|\alpha_{r}\right|^{2}=\exp\left[-\sum_{I}G^{\mathrm{A}}(z_{I};Z_{r})+\sum_{s\neq r}G^{\mathrm{A}}(Z_{r};Z_{s})+c\right]\,.
\]
With $\alpha_{r}$ fixed, $z_{I}\to Z_{r}$ implies that there exist
some punctures coming close to each other. %
}
\end{enumerate}
When the interaction points collide, $F_{N}^{(g)}$ could become singular
because $\partial^{2}\rho$'s at these points become $0$ and $T_{F}$'s
have singular OPE's. Since all the quantities which appear in $F_{N}^{(g)}$
can be expressed explicitly in terms of the theta functions defined
on $\Sigma_{N}^{(g)}$, possible singularities of $F_{N}^{(g)}$ also
originate from degenerations of the surface. The singularities of
$F_{N}^{(g)}$ arise only from these phenomena, because the world
sheet theory does not involve variables like superconformal ghost. 

Therefore, in order to study the possible divergences of the amplitude
$A_{N}^{(g)}$, we need to investigate the light-cone diagrams in
which 1 and/or 2 above happen. Light-cone diagrams with collisions
of interaction points are easily visualized. We shall study how light-cone
diagrams corresponding to degenerate Riemann surfaces look like, in
the next section.

\section{Light-cone diagrams in the degeneration limits\label{sec:Light-cone-diagrams-in}}

There are two types of degeneration, i.e. separating and nonseparating.
The expressions of various quantities in these limits are given in
\cite{Fay1973,Yamada:1980,Wentworth:1991,Wentworth:2008} from which
that of $\rho(z)$ can be obtained. The shape of the light-cone diagrams
can be deduced from the form of $\rho(z)$.

\subsection{Separating degeneration}

Let us first consider the separating degeneration in which a Riemann
surface $M$ degenerates into two surfaces $M_{1}$ and $M_{2}$ with
genera $g_{1}$ and $g_{2}$ respectively. We assume that $M$ corresponds
to a light-cone diagram with $N$ external lines, and $N_{1}$ of
them belong to $M_{1}$ and $N_{2}$ of them belong to $M_{2}$.

\subsubsection{$g_{1}g_{2}\protect\ne0$\label{sub:g1g2ne0}}

Let us first consider the case when both $g_{1}$ and $g_{2}$ are
positive. The degeneration can be described by a model $M_{t}$ constructed
as follows:
\begin{itemize}
\item Choose points $p_{j}\in M_{j}$ and a neighborhood $U_{j}$ of $p_{j}$
for $j=1,2$. Let $D$ be the unit disk in $\mathbb{C}$ and define
$z_{j}:U_{j}\to D$ to be the coordinate of $U_{j}$ such that $z_{j}(p_{j})=0$.
\item For $0<r<1$, let $rU_{j}$ be 
\[
rU_{j}\equiv\left\{ p\in U_{j};\ \left|z_{j}(p)\right|<r\right\} \,.
\]
For $t\in D$, glue together the surfaces $M_{j}/\left|t\right|U_{j}\ (j=1,2)$
by the identification 
\[
z_{1}z_{2}=t\,.
\]
The surface obtained is denoted by $M_{t}$. The degeneration limit
corresponds to $t\to0$. 
\end{itemize}
The complex coordinate $\rho_{t}(z)$ on the light-cone diagram corresponding
to $M_{t}$ is given by
\[
\rho_{t}(z)=\sum_{r=1}^{N}\alpha_{r}\left[\ln E_{t}(z,Z_{r})-2\pi i\int_{P_{0}}^{z}\omega_{t}\frac{1}{\mathop{\mathrm{Im}}\Omega_{t}}\mathop{\mathrm{Im}}\int_{P_{0}}^{Z_{r}}\omega_{t}\right]\,,
\]
where $E_{t}(z,w),\,\omega_{t},\,\Omega_{t}$ denote the prime form,
the canonical basis of the holomorphic abelian differentials and the
period matrix of $M_{t}$ respectively. The base point $P_{0}$ is
taken to be included in $M_{1}/\sqrt{\left|t\right|}U_{1}$. We take
the punctures $Z_{1},\cdots,Z_{N_{1}}$ to belong to $M_{1}/\sqrt{\left|t\right|}U_{1}$
and the punctures $Z_{N_{1}+1},\cdots,Z_{N}$ to belong to $M_{2}/\sqrt{\left|t\right|}U_{2}$.
Using the formulas of various quantities for $\left|t\right|\ll1$
given in \cite{Fay1973,Yamada:1980,Wentworth:1991,Wentworth:2008},
it is possible to show that $\rho_{t}(x_{1}),\rho_{t}(x_{2})$ for
$x_{j}\in M_{j}/\sqrt{\left|t\right|}U_{j}$ become 
\begin{eqnarray}
\rho_{t}(x_{1}) & = & \rho^{(1)}(x_{1})+\sum_{r_{2}=N_{1}+1}^{N}\alpha_{r_{2}}\ln E_{2}(p_{2},Z_{r_{2}})-\frac{\alpha_{p_{1}}}{2}\ln(-t)+\cdots\,,\nonumber \\
\rho_{t}(x_{2}) & = & \rho^{(2)}(x_{2})+\sum_{r_{1}=1}^{N_{1}}\alpha_{r_{1}}\ln E_{1}(p_{1},Z_{r_{1}})-\frac{\alpha_{p_{2}}}{2}\ln(-t)\nonumber \\
 &  & \quad+2\pi i\int_{P_{0}^{\prime}}^{p_{2}}\omega^{(2)}\frac{1}{\mathop{\mathrm{Im}}\Omega_{2}}\mathop{\mathrm{Im}}\left(\sum_{r_{2}=N_{1}+1}^{N}\alpha_{r_{2}}\int_{p_{2}}^{Z_{r_{2}}}\omega^{(2)}+\alpha_{p_{2}}\int_{P_{0}^{\prime}}^{p_{2}}\omega^{(2)}\right)\nonumber \\
 &  & \quad-2\pi i\int_{P_{0}}^{p_{1}}\omega^{(1)}\frac{1}{\mathop{\mathrm{Im}}\Omega_{1}}\mathop{\mathrm{Im}}\left(\sum_{r_{1}=1}^{N_{1}}\alpha_{r_{1}}\int_{P_{0}}^{Z_{r_{1}}}\omega^{(1)}+\alpha_{p_{1}}\int_{P_{0}}^{p_{1}}\omega^{(1)}\right)\nonumber \\
 &  & \quad+\cdots\,,\label{eq:separatingrho}
\end{eqnarray}
for $\left|t\right|\ll1$. Here the ellipses denote the terms higher
order in $t$ and 
\begin{eqnarray*}
\rho^{(1)}(x_{1}) & = & \sum_{r_{1}=1}^{N_{1}}\alpha_{r_{1}}\ln E_{1}(x_{1},Z_{r_{1}})+\alpha_{p_{1}}\ln E_{1}(x_{1},p_{1})\\
 &  & \quad-2\pi i\int_{P_{0}}^{x_{1}}\omega^{(1)}\frac{1}{\mathop{\mathrm{Im}}\Omega_{1}}\mathop{\mathrm{Im}}\left(\sum_{r_{1}=1}^{N_{1}}\alpha_{r_{1}}\int_{P_{0}}^{Z_{r_{1}}}\omega^{(1)}+\alpha_{p_{1}}\int_{P_{0}}^{p_{1}}\omega^{(1)}\right)\,,\\
\rho^{(2)}(x_{2}) & = & \sum_{r_{2}=N_{1}+1}^{N}\alpha_{r_{2}}\ln E_{2}(x_{2},Z_{r_{2}})+\alpha_{p_{2}}\ln E_{2}(x_{2},p_{2})\\
 &  & \quad-2\pi i\int_{P_{0}^{\prime}}^{x_{2}}\omega^{(2)}\frac{1}{\mathop{\mathrm{Im}}\Omega_{2}}\mathop{\mathrm{Im}}\left(\sum_{r_{2}=N_{1}+1}^{N}\alpha_{r_{2}}\int_{P_{0}^{\prime}}^{Z_{r_{2}}}\omega^{(2)}+\alpha_{p_{2}}\int_{P_{0}^{\prime}}^{p_{2}}\omega^{(2)}\right)\,,\\
\alpha_{p_{1}} & = & \sum_{r_{2}=N_{1}+1}^{N}\alpha_{r_{2}}=-\alpha_{p_{2}}=-\sum_{r_{1}=1}^{N_{1}}\alpha_{r_{1}}\,,
\end{eqnarray*}
with $P_{0}^{\prime}\in M_{2}/\sqrt{\left|t\right|}U_{2}$. $E_{j}(z,w),\,\omega^{(j)},\,\Omega_{j}$
denote the prime form, the canonical basis of the holomorphic abelian
differentials and the period matrix of $M_{j}\,(j=1,2)$. 

\begin{figure}
\begin{centering}
\includegraphics[scale=0.5]{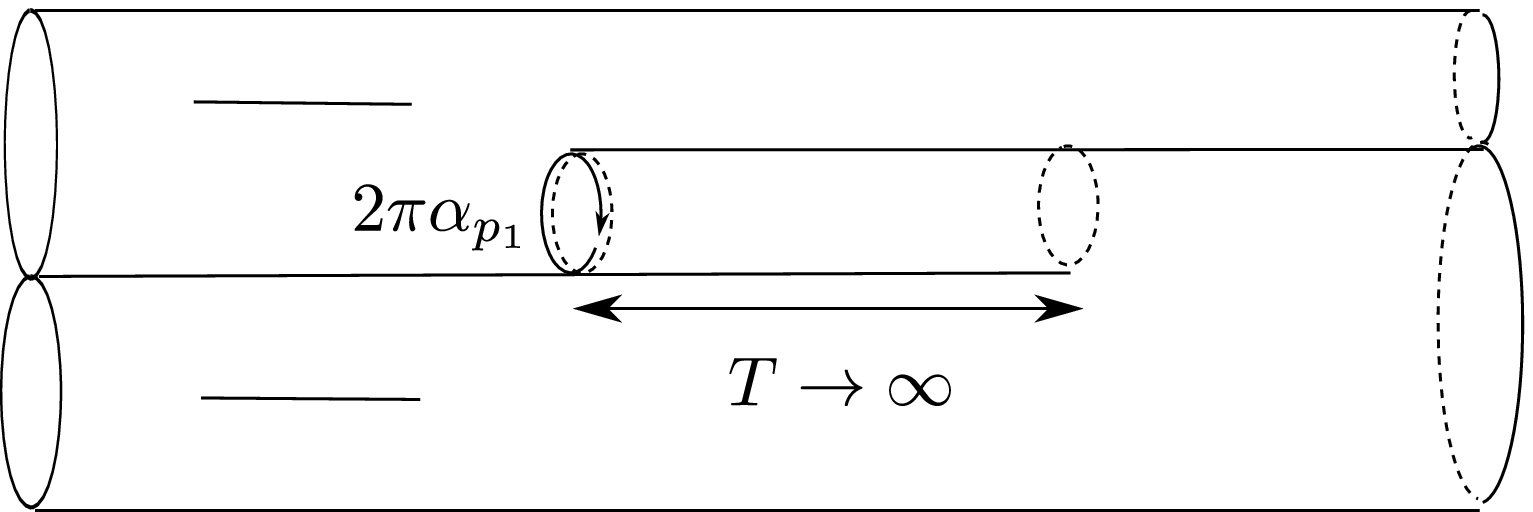}
\par\end{centering}

\protect\caption{\label{fig:separating1}Light-cone diagram corresponding to a separating
degeneration with $\alpha_{p_{1}}\protect\ne0$.}
\end{figure}

From (\ref{eq:separatingrho}), we can see how the light-cone diagrams
corresponding to the separating degeneration should look like. They
are classified according to the values of $\alpha_{p_{1}},N_{1},N_{2}$
as follows:
\begin{itemize}
\item $\alpha_{p_{1}}\ne0$\\
If $\alpha_{p_{1}}\ne0$, $\rho^{(1)},\rho^{(2)}$ can be considered
as the coordinates defined on light-cone diagrams with $N^{(1)}+1,\, N^{(2)}+1$
external lines respectively. (\ref{eq:separatingrho}) implies that
the limit $t\to0$ corresponds to the one in which the length of an
internal line with $\alpha=\alpha_{p_{1}}$ becomes infinite in the
light-cone diagram. A light-cone diagram of this type is presented
in Fig. \ref{fig:separating1}. $M_{1}$ and $M_{2}$ correspond to
light-cone diagrams with the coordinates $\rho^{(1)}$ and $\rho^{(2)}$
respectively. 
\end{itemize}

\begin{figure}
\begin{centering}
\includegraphics[scale=0.5]{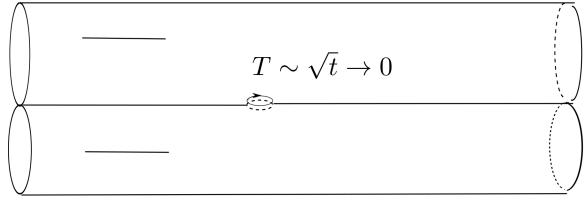}
\par\end{centering}

\protect\caption{\label{fig:separating2}Light-cone diagram corresponding to a separating
degeneration with $\alpha_{p_{1}}=0$.}
\end{figure}

\begin{itemize}
\item $\alpha_{p_{1}}=0,\, N_{1}N_{2}\ne0$\\
When $\alpha_{p_{1}}=0,\, N_{1}N_{2}\ne0$, higher order terms in
(\ref{eq:separatingrho}) with respect to $t$ become important. We
get 
\begin{eqnarray}
\partial\rho_{t}(x_{1}) & \sim & \partial\rho^{(1)}(x_{1})-t\partial_{p_{1}}\partial_{x_{1}}\ln E_{1}(x_{1},p_{1})\partial\rho^{(2)}(p_{2})+\cdots\,,\label{eq:rhot1}\\
\partial\rho_{t}(x_{2}) & \sim & \partial\rho^{(2)}(x_{2})-t\partial_{p_{2}}\partial_{x_{2}}\ln E_{2}(x_{2},p_{2})\partial\rho^{(1)}(p_{1})+\cdots\,,\label{eq:rhot2}
\end{eqnarray}
where $\rho^{(1)},\rho^{(2)}$ in this case are given by those above
with $\alpha_{p_{1}}=\alpha_{p_{2}}=0$. Since $N_{1}N_{2}\ne0$,
neither of $\rho^{(1)}$ and $\rho^{(2)}$ is identically $0$. For
$x_{1}\sim p_{1},\, x_{2}\sim p_{2}$, the coordinates $z_{1}=z_{1}(x_{1}),\, z_{2}=z_{2}(x_{2})$
can be used to describe the region and we get 
\begin{eqnarray}
\partial\rho_{t}(z_{1}) & = & c_{1}-\frac{c_{2}t}{z_{1}^{2}}+\cdots\,,\label{eq:drhot1}\\
\partial\rho_{t}(z_{2}) & = & c_{2}-\frac{c_{1}t}{z_{2}^{2}}+\cdots\,,\label{eq:drhot2}
\end{eqnarray}
where 
\begin{eqnarray*}
c_{1} & = & \left.\partial\rho^{(1)}(z_{1})\right|_{z_{1}=0}\,,\\
c_{2} & = & \left.\partial\rho^{(2)}(z_{2})\right|_{z_{2}=0}\,.
\end{eqnarray*}
Defining 
\begin{equation}
z\equiv\sqrt{t}z_{1}\,,\label{eq:zroottz1}
\end{equation}
which is a good coordinate of the region $z_{1}\sim z_{2}\sim\sqrt{t}$,
\[
\rho_{t}(z)\sim\mbox{constant}+\sqrt{t}(c_{1}z+\frac{c_{2}}{z})\,.
\]
When $c_{1}c_{2}\ne0$, the degeneration of this type can be represented
by the light-cone diagram depicted in Fig. \ref{fig:separating2}.
There are two interaction points in the light-cone diagram corresponding
to $M_{t}$ which come close to each other and the surface develops
a narrow neck in the limit $t\to0$. They are included in a region
which has coordinate size of order $\sqrt{t}$ in the light-cone diagram
and shrinks to a point in the limit $t\to0$. The case $c_{1}c_{2}=0$
corresponds to the case where some of the interaction points on $M_{1},M_{2}$
come close to $p_{1},p_{2}$ respectively.
\end{itemize}

\begin{figure}
\begin{centering}
\includegraphics[scale=0.5]{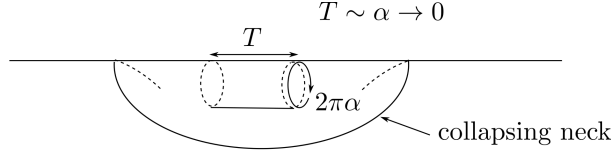}
\par\end{centering}

\protect\caption{\label{fig:separating3}Light-cone diagram corresponding to a separating
degeneration with $N_{2}=0$.}
\end{figure}

\begin{itemize}
\item $N_{1}N_{2}=0$\\
When $N_{2}=0$ for example, (\ref{eq:rhot1}) and (\ref{eq:rhot2})
become
\begin{eqnarray}
\partial\rho_{t}(x_{1}) & \sim & \partial\rho^{(1)}(x_{1})+\cdots\,,\nonumber \\
\partial\rho_{t}(x_{2}) & \sim & -t\partial_{p_{2}}\partial_{x_{2}}\ln E_{2}(x_{2},p_{2})\partial\rho^{(1)}(p_{1})+\cdots\,.\label{eq:specialrho23}
\end{eqnarray}
Therefore $M_{2}/\sqrt{\left|t\right|}U_{2}$ corresponds to a tiny
region in the light-cone diagram corresponding to $M_{t}$, which
shrinks to a point in the limit $t\to0$ and the collapsing neck has
the coordinate size of order $t$ on the light-cone diagram. An example
of such a situation is shown in Fig. \ref{fig:separating3}. $\partial\rho^{(1)}(p_{1})=0$
in this case can be regarded as the situation in which some of the
interaction points on $M_{1}$ come close to $p_{1}$.
\end{itemize}

\subsubsection{$g_{1}g_{2}=0$}

The case where $g_{1}$ or $g_{2}$ vanish corresponds to the situation
in which some of the punctures $Z_{1},\cdots,Z_{N}$ come close to
each other. Let us consider the separating degeneration in which a
Riemann surface $M$ degenerates into two surfaces $M_{1}$ and $M_{2}$
with genera $g_{1},g_{2}=0$ respectively. We assume that the punctures
$Z_{1},\cdots,Z_{N_{1}}$ belong to $M_{1}$ and $Z_{N_{1}+1},\cdots,Z_{N}$
belong to $M_{2}$. Such a degeneration can be described as follows.
Choose a point $p\in M$ and a neighborhood $U$ of $p$ such that
the punctures $Z_{N_{1}+1},\cdots,Z_{N}$ are included in $U$. Let
$D$ be the unit disk in $\mathbb{C}$ and define $z_{1}:U\to D$
to be the coordinate of $U$ such that $z_{1}(p)=0$. We take 
\[
z_{1}(Z_{r_{2}})=tz_{r_{2}}\,(r_{2}=N_{1}+1,\cdots,N)\,,
\]
and consider the limit $t\to0$ with $z_{r_{2}}$ fixed. When $\left|t\right|\ll1$,
$\rho(x_{1})$ for $x_{1}\in M/\sqrt{\left|t\right|}U$ becomes
\begin{eqnarray*}
\rho(x_{1}) & \sim & \sum_{r_{1}=1}^{N_{1}}\alpha_{r_{1}}\ln E(x_{1},Z_{r_{1}})+\alpha_{p_{1}}\ln E(x_{1},p_{1})\\
 &  & \quad-2\pi i\int_{P_{0}}^{x_{1}}\omega\frac{1}{\mathop{\mathrm{Im}}\Omega}\mathop{\mathrm{Im}}\left(\sum_{r_{1}=1}^{N_{1}}\alpha_{r_{1}}\int_{P_{0}}^{Z_{r_{1}}}\omega^{(1)}+\alpha_{p_{1}}\int_{P_{0}}^{p_{1}}\omega^{(1)}\right)\\
 & \equiv & \rho^{(1)}(x_{1})\,.
\end{eqnarray*}
For $x_{2}\in U$ such that $z_{1}(x_{2})=\mathcal{O}(\left|t\right|)$,
defining $z_{2}\equiv\frac{z_{1}(x_{2})}{t}$, 
\begin{eqnarray*}
\rho(x_{2}) & \sim & \rho^{(2)}(z_{2})-\frac{\alpha_{p_{2}}}{2}\ln t+\lim_{x\to p_{1}}(\rho^{(1)}(x)-\alpha_{p_{1}}\ln(x-p_{1}))\,,\\
\rho^{(2)}(z_{2}) & = & \sum_{r_{2}=N_{1}+1}^{N}\alpha_{r_{2}}\ln(z_{2}-z_{r_{2}})\,,
\end{eqnarray*}
where 
\[
\alpha_{p_{1}}=\sum_{r_{2}=N_{1}+1}^{N}\alpha_{r_{2}}=-\alpha_{p_{2}}=-\sum_{r_{1}=1}^{N_{1}}\alpha_{r_{1}}\,.
\]

If $\alpha_{p_{1}}\ne0$, these formulas suggest that the limit corresponds
to the one in which the length of an internal line with circumference
$2\pi\alpha_{p_{1}}$ becomes infinite in the light-cone diagram.
If $\alpha_{p_{1}}=0$, the collapsing neck can be described by a
local coordinate $z\equiv\sqrt{t}z_{1}$ and 
\[
\rho(z)\sim\mbox{constant}+\sqrt{t}(c_{1}z+\frac{c_{2}}{z})\,,
\]
where 
\begin{eqnarray*}
c_{1} & = & \left.\partial\rho^{(1)}(z_{1})\right|_{z_{1}=0}\,,\\
c_{2} & = & \left.-z_{2}^{2}\partial\rho^{(2)}(z_{2})\right|_{z_{2}=\infty}\,.
\end{eqnarray*}
Namely the light-cone diagrams are locally the same as those we encountered
in the case where $g_{1},\, g_{2}$ are both positive.

\subsection{Nonseparating degeneration}

Next let us consider the nonseparating degeneration in which a Riemann
surface of genus $g+1$ degenerates into a surface $M$ of genus $g$.
The degeneration can be described by a model $M_{t}$ constructed
as follows:
\begin{itemize}
\item Choose points $p_{1},p_{2}\in M$ and their disjoint neighborhoods
$U_{1},U_{2}$. Let $z_{j}:U_{j}\to D$ be the coordinate of $U_{j}$
such that $z_{j}(p_{j})=0$.
\item For $t\in D$, glue together the surfaces $M_{j}/\left|t\right|U_{j}\ (j=1,2)$
by the identification 
\[
z_{1}z_{2}=t\,.
\]
The surface obtained is denoted by $M_{t}$. The degeneration limit
corresponds to $t\to0$. 
\end{itemize}
The coordinate $\rho_{t}(z)$ on the light-cone diagram corresponding
to $M_{t}$ is given by
\[
\rho_{t}(z)=\sum_{r=1}^{N}\alpha_{r}\left[\ln E_{t}(z,Z_{r})-2\pi i\int_{P_{0}}^{z}\omega_{t}\frac{1}{\mathop{\mathrm{Im}}\Omega_{t}}\mathop{\mathrm{Im}}\int_{P_{0}}^{Z_{r}}\omega_{t}\right]\,.
\]
 Using the formulas given in \cite{Fay1973,Yamada:1980,Wentworth:1991,Wentworth:2008},
it is straightforward to deduce that $\rho_{t}(z)$ can be expressed
as
\begin{eqnarray}
\rho_{t}(z) & = & \rho(z)\nonumber \\
 &  & +\alpha_{p_{1}}\left[\ln E(z,p_{1})-2\pi i\int_{P_{0}}^{z}\omega\frac{1}{\mathop{\mathrm{Im}}\Omega}\mathop{\mathrm{Im}}\int_{P_{0}}^{p_{1}}\omega\right]\nonumber \\
 &  & +\alpha_{p_{2}}\left[\ln E(z,p_{2})-2\pi i\int_{P_{0}}^{z}\omega\frac{1}{\mathop{\mathrm{Im}}\Omega}\mathop{\mathrm{Im}}\int_{P_{0}}^{p_{2}}\omega\right]\,,\nonumber \\
 &  & +\mathrm{constant}+\mathcal{O}(t)\,,\label{eq:nonseparatingrho}
\end{eqnarray}
for $\left|t\right|\ll1$. Here 
\[
\rho(z)=\sum_{r=1}^{N}\alpha_{r}\left[\ln E(z,Z_{r})-2\pi i\int_{P_{0}}^{z}\omega\frac{1}{\mathop{\mathrm{Im}}\Omega}\mathop{\mathrm{Im}}\int_{P_{0}}^{Z_{r}}\omega\right]
\]
 gives the coordinate of the light-cone diagram corresponding to $M$
and 
\[
\alpha_{p_{2}}=-\alpha_{p_{1}}=\frac{1}{2\pi}\mathrm{Re}(\rho(p_{2})-\rho(p_{1}))\frac{1-\frac{2\pi}{\ln\left|t^{\prime}\right|}\mathop{\mathrm{Im}}\int_{p_{1}}^{p_{2}}\omega\frac{1}{\mathop{\mathrm{Im}}\Omega}\mathop{\mathrm{Im}}\int_{p_{1}}^{p_{2}}\omega}{-\frac{\ln\left|t^{\prime}\right|}{2\pi}+\frac{2\pi}{\ln\left|t^{\prime}\right|}\left(\mathop{\mathrm{Im}}\int_{p_{1}}^{p_{2}}\omega\frac{1}{\mathop{\mathrm{Im}}\Omega}\mathop{\mathrm{Im}}\int_{p_{1}}^{p_{2}}\omega\right)^{2}}\,,
\]
with
\[
t^{\prime}=\frac{t}{E(p_{1},p_{2})E(p_{2},p_{1})}\,.
\]

\begin{figure}
\begin{centering}
\includegraphics[scale=0.5]{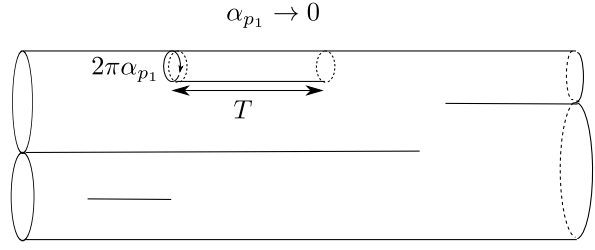}
\par\end{centering}

\protect\caption{\label{fig:nonseparating1}Nonseparating degeneration with $\mathrm{Re}(\rho(p_{2})-\rho(p_{1}))\protect\ne0$.}
\end{figure}

Using (\ref{eq:nonseparatingrho}), we can classify the light-cone
diagrams corresponding to the nonseparating degeneration according
to the value of $\mathrm{Re}(\rho(p_{2})-\rho(p_{1}))$ as follows%
\footnote{Here we assume that $\mathrm{Re}(\rho(p_{2})-\rho(p_{1}))$ is fixed
in the limit $t\to0$.%
}:
\begin{itemize}
\item $\mathrm{Re}(\rho(p_{2})-\rho(p_{1}))\ne0$\\
If $\mathrm{Re}(\rho(p_{2})-\rho(p_{1}))\ne0$, we can see from (\ref{eq:nonseparatingrho})
that this kind of degeneration corresponds to a limit of light-cone
diagram in which the circumference of an internal line tends to zero
as depicted in Fig. \ref{fig:nonseparating1}.
\end{itemize}

\begin{figure}
\begin{centering}
\includegraphics[scale=0.5]{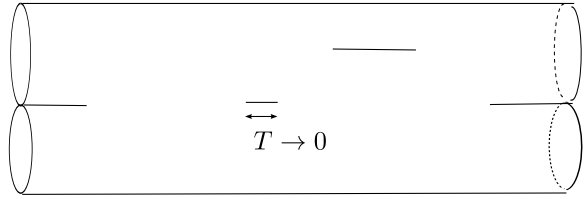}
\par\end{centering}

\protect\caption{\label{fig:nonseparating2}An example of nonseparating degeneration
with $\mathrm{Re}(\rho(p_{2})-\rho(p_{1}))=0$.}
\end{figure}

\begin{figure}
\begin{centering}
\includegraphics[scale=0.5]{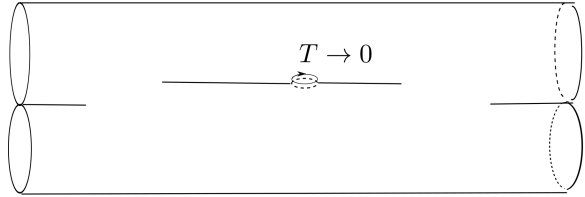}
\par\end{centering}

\protect\caption{\label{fig:nonseparating3}Another example of nonseparating degeneration
with $\mathrm{Re}(\rho(p_{2})-\rho(p_{1}))=0$.}
\end{figure}

\begin{itemize}
\item $\mathrm{Re}(\rho(p_{2})-\rho(p_{1}))=0$\\
When $\mathrm{Re}(\rho(p_{2})-\rho(p_{1}))=0$, $\rho_{t}(x)$ becomes
\[
\rho_{t}(x)=\rho(x)+t\left[\partial_{p_{1}}\ln E(x,p_{1})\partial\rho(p_{2})+\partial_{p_{2}}\ln E(x,p_{2})\partial\rho(p_{1})\right]+\cdots\,.
\]
Therefore, for $x\sim p_{1}$
\begin{equation}
\rho_{t}(x)=\mbox{constant}+c_{1}z_{1}+\frac{c_{2}t}{z_{1}}+\cdots\,,\label{eq:specialrho15-1}
\end{equation}
where $z_{1}=z_{1}(x)$ and 
\begin{eqnarray*}
c_{1} & = & \left.\partial\rho(z_{1})\right|_{z_{1}=0}\,,\\
c_{2} & = & \left.\partial\rho(z_{2})\right|_{z_{2}=0}\,.
\end{eqnarray*}
Similarly, for $x\sim p_{2}$
\begin{equation}
\rho_{t}(x)=\mbox{constant}+c_{2}z_{2}+\frac{c_{1}t}{z_{2}}+\cdots\,,\label{eq:specialrho25-1}
\end{equation}
where $z_{2}=z_{2}(x)$. Hence the $\rho_{t}$ has the same expression
as those in (\ref{eq:drhot1}), (\ref{eq:drhot2}). When $c_{1}c_{2}\ne0$,
the degeneration of this type can be represented by the light-cone
diagrams depicted in Figs. \ref{fig:nonseparating2} and \ref{fig:nonseparating3}.
There are two interaction points in the light-cone diagram corresponding
to $M_{t}$ which come close to each other and the surface develops
a narrow neck in the limit $t\to0$. They are included in a region
which has coordinate size of order $\sqrt{t}$ on the light-cone diagram
and shrinks to a point in the limit $t\to0$. $c_{1}c_{2}=0$ corresponds
to the case in which some interaction points on $M$ come close to
$p_{1}$ or $p_{2}$.
\end{itemize}

\subsection{Combined limits\label{sub:Combined-limits-1}}

In the discussions above, we have implicitly assumed that the parameters
$\alpha_{p_{1}}=\sum\alpha_{r_{2}}$ or $\mathrm{Re}(\rho(p_{2})-\rho(p_{1}))$
are fixed in taking the degeneration limit $t\to0$. This is true
if the degeneration considered is the only one which occurs on the
surface. If we consider the situation where several degenerations
happen simultaneously, these parameters are not necessarily fixed
and we encounter new classes of light-cone diagrams corresponding
to degeneration. 

\begin{figure}
\begin{centering}
\includegraphics[scale=0.5]{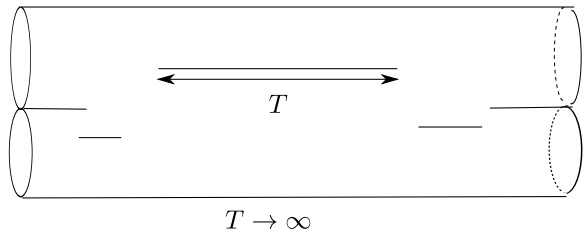}
\par\end{centering}

\protect\caption{\label{fig:combination}A combination of degeneration limits. }
\end{figure}

Taking such situations into account, the light-cone diagram in the
degeneration limits are classified by the behavior of the parameter
$\alpha_{p_{1}}=-\alpha_{p_{2}}$ in the limit $t\to0$. For $z_{1},z_{2}\ll1,\, t\ll1$,
we have 
\begin{eqnarray*}
\partial\rho_{t}(z_{1}) & \sim & \frac{\alpha_{p_{1}}}{z_{1}}+\mbox{higher order terms in }t\,,\\
\partial\rho_{t}(z_{2}) & \sim & \frac{\alpha_{p_{2}}}{z_{2}}+\mbox{higher order terms in }t\,,
\end{eqnarray*}
irrespective of whether the degeneration is separating or nonseparating.
If $\alpha_{p_{1}}$tends to a finite nonvanishing value in the limit
$t\to0$, these imply that the light-cone diagram develops an infinitely
long cylinder with finite width. The diagrams depicted in Figs. \ref{fig:separating1},
\ref{fig:combination} are in this class. If $\alpha_{p_{1}}$tends
to $0$ as $t\to0$, the light-cone diagram develops a cylinder with
vanishing width, provided $\frac{\alpha_{p_{1}}}{z_{1}},\frac{\alpha_{p_{2}}}{z_{2}}$
dominate the higher order terms in $t$. For example, if $\partial\rho_{1}(p_{1})\partial\rho_{2}(p_{2})\ne0$,
defining $z=t^{\frac{1}{2}}z_{1}$ as in (\ref{eq:zroottz1}), we
have
\begin{equation}
\rho_{t}(z)\sim\alpha_{p_{1}}\ln z+t^{\frac{1}{2}}(c_{1}z+\frac{c_{2}}{z})+\cdots\,.\label{eq:log}
\end{equation}
Therefore if $\alpha_{p_{1}}$ goes to $0$ slower than $t^{\frac{1}{2}}$
in the limit $t\to0$, the surface develops a cylinder with vanishing
width. If $t^{-\frac{1}{2}}\alpha_{p_{1}}$ tends to a finite value
in the limit $t\to0$, we have 
\[
\rho_{t}(z)\sim t^{\frac{1}{2}}(c_{1}z+\frac{c_{2}}{z}+\alpha\ln z)+\mbox{constant}\,,
\]
where $\alpha=\lim_{t\to0}t^{-\frac{1}{2}}\alpha_{p_{1}}$. The degeneration
of this type can be represented by the light-cone diagrams depicted
in Figs. \ref{fig:separating2}, \ref{fig:nonseparating2}, \ref{fig:nonseparating3},
but this time 
\[
\oint_{\mbox{neck}}dz\partial\rho_{t}\sim2\pi it^{\frac{1}{2}}\alpha\ne0\,,
\]
namely the coordinate $\rho$ can be multivalued around the neck.
If $\alpha_{p_{1}}$ goes to $0$ faster than $t^{\frac{1}{2}}$,
the first term in (\ref{eq:log}) can be ignored and we have $\rho$
single-valued around the neck. $\partial\rho_{1}(p_{1})\partial\rho_{2}(p_{2})=0$
case corresponds to the situation in which some interaction points
come close to $p_{1},\, p_{2}$.

\subsection{Classification of the light-cone diagrams in the degeneration limits
\label{sub:Classification-of-the}}

To summarize, what happens to a light-cone diagram in the degeneration
limit can be classified by the behavior of the degenerating cycle
as follows:
\begin{enumerate}
\item The light-cone diagram develops an infinitely long cylinder with nonvanishing
width. \\
The diagrams presented in Figs. \ref{fig:separating1}, \ref{fig:combination}
belong to this class. 
\item The light-cone diagram develops an infinitely thin cylinder. \\
The diagram depicted in Fig. \ref{fig:nonseparating1} belongs to
this class. 
\item The light-cone diagram develops a narrow neck included in a region
which shrinks to a point. \\
The diagrams shown in Figs. \ref{fig:separating2}, \ref{fig:separating3},
\ref{fig:nonseparating2}, \ref{fig:nonseparating3} belong to this
class. 
\end{enumerate}
A diagram of the first type includes a long cylinder and can definitely
be considered as corresponding to the infrared region of the integration
over the moduli space. The divergent contributions from such diagrams
can be made finite by the Feynman $i\varepsilon$ as will be shown
in the next section. On the other hand, that of the third type appears
to correspond to the ultraviolet region with respect to the world
sheet metric $ds^{2}=d\rho d\bar{\rho}$, although the collapsing
neck is conformally equivalent to a long cylinder. We need to introduce
a linear dilaton background to regularize the divergences coming from
the diagrams in this category and the Feynman $i\varepsilon$ plays
no roles in this case. The second type is something in between and
we need both $i\varepsilon$ and a linear dilaton background to deal
with the divergence coming from such a configuration.

\section{Divergences of the amplitudes\label{sec:Degenerate-Riemann-surfaces}}

Divergences of the amplitudes arise from the diagrams in which degenerations
and/or collisions of interaction points occur. In this section, we
examine the divergences of the amplitude (\ref{eq:ANlin}) corresponding
to these configurations.

\subsection{Diagrams which involve cylinders with infinite length and nonvanishing
width}

\begin{figure}
\begin{centering}
\includegraphics{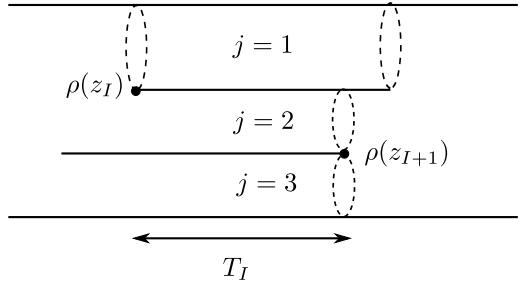}
\par\end{centering}

\protect\caption{Cylinders in a light-cone diagram.\label{fig:Cylinders-in-a}}
\end{figure}

Let us first consider the first type of degeneration limit
in the classification in subsection
\ref{sub:Classification-of-the}, in which
the diagram develops a cylinder with height $T\to\infty$ and nonvanishing
circumference. In the light-cone gauge perturbation theory, such cylinders
appear as follows. Let us order the interaction points $z_{I}\,(I=1,\ldots,2g-2+N)$
so that 
\[
\mathrm{Re}\rho(z_{1})\leq\mathrm{Re}\rho(z_{2})\leq\cdots\leq\mathrm{Re}\rho(z_{2g-2+N})\,,
\]
and define the moduli parameters corresponding to the heights as
\[
T_{I}\equiv\mathrm{Re}\rho(z_{I+1})-\mathrm{Re}\rho(z_{I})\,(I=1,\ldots,2g-3+N)\,.
\]
Long cylinders appear in the limit $T_{I}\to\infty$. In studying
the divergence from this region of the moduli space, the relevant
part of the amplitude is of the form 
\begin{equation}
\int_{0}^{\infty}dT_{I}\exp\left[-T_{I}\left(\sum_{j}\frac{L_{0}^{(j)}+\bar{L}_{0}^{(j)}-1+Q^{2}-i\varepsilon}{\alpha_{j}}-\sideset{}{^{\prime}}\sum_{r}p_{r}^{-}\right)\right]\,,\label{eq:TIint}
\end{equation}
where $j$ labels the cylinders which include the region $\mathrm{Re}\rho(z_{I})\leq\mathrm{Re}\rho\leq\mathrm{Re}\rho(z_{I+1})$
(see Fig. \ref{fig:Cylinders-in-a}) and $\alpha_{j},\, L_{0}^{(j)},\,\bar{L}_{0}^{(j)}$
denote the $\alpha,\, L_{0},\,\bar{L}_{0}$ defined on the $j$-th
cylinder respectively. $r$ labels the external lines and $\sum'$
denotes the sum over those with $p_{r}^{-}>0$. The degeneration limit
corresponds to the one in which $T_{I}$ goes to infinity. Since the
lowest eigenvalue of $L_{0}^{(j)}+\bar{L}_{0}^{(j)}$ is $1-Q^{2}$
for the GSO even sector, the integral (\ref{eq:TIint}) could diverge
because of the contribution from the limit. 

The divergence coming from this kind of degeneration can be dealt
with by deforming the contour of integration over $T_{I}$ \cite{Witten2015,Berera1994a,Mandelstam:2008fa}.
As is suggested in \cite{Witten2015}, we take the contour to be as
\begin{equation}
\left(\int_{0}^{T_{0}}+\int_{T_{0}}^{T_{0}+i\infty}\right)dT_{I}\,,\label{eq:decTI}
\end{equation}
with $T_{0}\gg1$. Since
\begin{eqnarray*}
 &  & \left|\exp\left[-(T_{0}+ia)\left(\sum_{j}\frac{L_{0}^{(j)}+\bar{L}_{0}^{(j)}-1+Q^{2}-i\varepsilon}{\alpha_{j}}-\sideset{}{^{\prime}}\sum_{r}p_{r}^{-}\right)\right]\right|\\
 &  & \qquad=\left|\exp\left[-T_{0}\left(\sum_{j}\frac{L_{0}^{(j)}+\bar{L}_{0}^{(j)}-1+Q^{2}-i\varepsilon}{\alpha_{j}}-\sideset{}{^{\prime}}\sum_{r}p_{r}^{-}\right)\right]\right|\exp\left(-a\varepsilon\sum_{j}\frac{1}{\alpha_{j}}\right)\,,
\end{eqnarray*}
the second integral in (\ref{eq:decTI}) yields a finite result for
$\varepsilon>0$. Thus the integral over $T_{I}$ is essentially cut
off at $T_{I}=T_{0}$ and the degeneration becomes harmless. 

In order to get the amplitudes, we need to take the limit $\varepsilon\to0$.
Notice that the divergences associated with the tadpole graphs correspond
to the separating degeneration with $N_{1}N_{2}=0$ described in subsection
\ref{sub:g1g2ne0}. As we will see in the following subsections, they
are regularized by taking $Q^{2}$ large enough and the Feynman $i\varepsilon$
is irrelevant. The modified momentum conservation law of $p^{1}$
for $g$-loop two point function is given by
\[
p_{1}^{1}+p_{2}^{1}+2Q(1-g)=0\,.
\]
Therefore if $p_{1}^{\mu}$ is on-shell, $p_{2}^{\mu}$ is generically
off-shell, for $g\ne0$. This implies that the divergences associated
with mass renormalization are also regularized by taking $Q\ne0$.
Therefore the procedures given in \cite{Witten:2012bh,Witten:2013cia,Pius2014a,Pius2014c}
shall be relevant in taking the limit $Q\to0$ rather than $\varepsilon\to0$.
Possible divergences in the limit $\varepsilon\to0$ can be analyzed
as in the usual field theory and we expect that they cancel each other
if one calculates physical quantities.

\subsection{Singular behavior of $F_{N}^{(g)}$\label{sub:Singular-behavior-of-2}}

Since the first type of degeneration in the classification in subsection
\ref{sub:Classification-of-the} is taken care of, what we should
grapple with are other types of singularities, namely the degenerations
of types 2 and 3 and the collisions of interaction points. The integration
variables in the expression (\ref{eq:ANlin}) are given by differences
of the coordinates $\rho,\bar{\rho}$ of the interaction points and
magnitudes of jump discontinuities of $\rho,\bar{\rho}$. Let $x^{j}\in\mathbb{R}\,(j=1,\cdots,n)$
denote these integration variables. The calculation of the amplitude
boils down to that of an integral
\begin{equation}
\int d^{n}xF_{N}^{(g)}(\vec{x})\,,\label{eq:intf}
\end{equation}
where $F_{N}^{(g)}(\vec{x})$ denotes the $F_{N}^{(g)}$ as a function
of the variables $x^{1},\cdots,x^{n}$. The singularities we are dealing
with occur when interaction points collide and/or cylinders become
infinitely thin. Therefore a necessary and sufficient condition for
$\vec{x}$ to correspond to such singularities can be expressed as
\begin{equation}
\vec{v}_{k}\cdot\vec{x}=0\,,\label{eq:vi}
\end{equation}
for some $\vec{v}_{k}\in\mathbb{R}^{n}\,(k=1,\cdots)$. In order to
study the behavior of $F_{N}^{(g)}(\vec{x})$ at these singularities,
it is convenient to express $F_{N}^{(g)}(\vec{x})$ as
\[
F_{N}^{(g)}(\vec{x})=e^{-\frac{1-Q^{2}}{2}\Gamma\left[\sigma;g_{z\bar{z}}^{\mathrm{A}}\right]}\left\langle \mathcal{O}_{N}^{(g)}\right\rangle \,,
\]
where 
\begin{eqnarray*}
\left\langle \mathcal{O}\right\rangle  & \equiv & \int\left[dX^{i}d\psi^{i}d\bar{\psi}^{i}\right]_{g_{z\bar{z}}^{\mathrm{A}}}e^{-S\left[X^{i},\psi^{i},\bar{\psi}^{i};g_{z\bar{z}}^{\mathrm{A}}\right]}\mathcal{O}\,,\\
\mathcal{O}_{N}^{(g)} & \equiv & \prod_{I=1}^{2g-2+N}\left(\left|\partial^{2}\rho\left(z_{I}\right)\right|^{-\frac{3}{2}}T_{F}^{\mathrm{LC}}\left(z_{I}\right)\bar{T}_{F}^{\mathrm{LC}}\left(\bar{z}_{I}\right)\right)\prod_{r=1}^{N}V_{r}^{\mathrm{LC}}\,.
\end{eqnarray*}

The integrals over the lengths of cylinders are essentially cut off
by taking $\varepsilon>0$. Therefore we should worry about the singularity
of $F_{N}^{(g)}(\vec{x})$ at finite values of the coordinates $x^{j}$.
Such singularities can be studied by calculating the derivatives of
$F_{N}^{(g)}(\vec{x})$ with respect to $x^{1},\cdots,x^{n}$, which
can be expressed by contour integrals of the correlation functions 
with energy-momentum tensor insertions:
\[
e^{-\frac{1-Q^{2}}{2}\Gamma\left[\sigma;g_{z\bar{z}}^{\mathrm{A}}\right]}\left\langle T(\rho)\mathcal{O}_{N}^{(g)}\right\rangle ,\, e^{-\frac{1-Q^{2}}{2}\Gamma\left[\sigma;g_{z\bar{z}}^{\mathrm{A}}\right]}\left\langle \bar{T}(\bar{\rho})\mathcal{O}_{N}^{(g)}\right\rangle \,.
\]
 As we will see in the following, these correlation functions become
singular when $\rho,\bar{\rho}$ coincide with those of the interaction
points. Therefore the contour integrals diverge when the interaction
points pinch the contour. This is exactly what happens in the situations
we are dealing with. Away from the singularities, $F_{N}^{(g)}(\vec{x})$
is a differentiable function of the parameters $x^{j}$.

\subsubsection{Singular behavior of $F_{N}^{(g)}(\vec{x})$ associated with the
configuration depicted in Fig. \ref{fig:separating3}\label{sub:Singular-behavior-of-1}}

\begin{figure}
\begin{centering}
\includegraphics[scale=0.5]{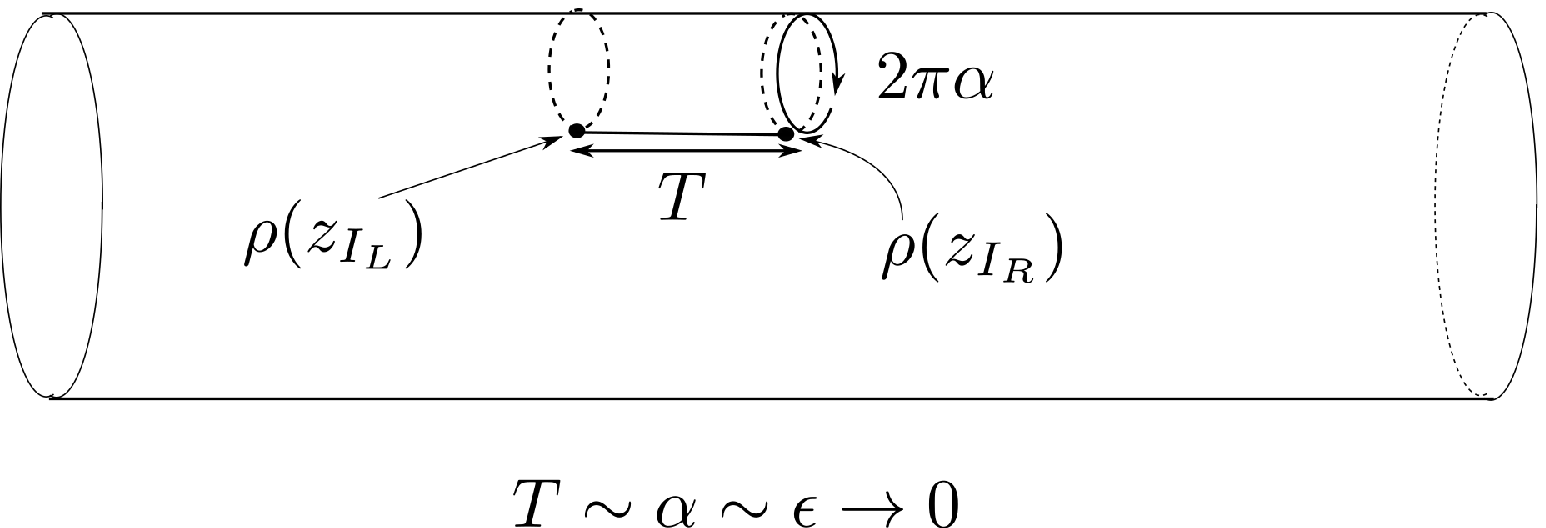}
\par\end{centering}

\protect\caption{\label{fig:scaling1}Fig. \ref{fig:separating3} embedded in a light-cone
diagram. }
\end{figure}

As an example, let us study the singular behavior of $F_{N}^{(g)}(\vec{x})$
in the limit illustrated in Fig. \ref{fig:separating3}. We here consider
the situation where the tiny cylinder is embedded in a light-cone
diagram as depicted in Fig. \ref{fig:scaling1}. 

\begin{figure}
\begin{centering}
\includegraphics{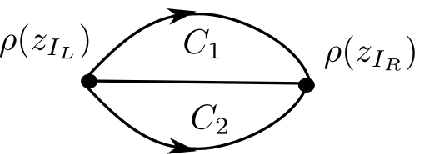}
\par\end{centering}

\protect\caption{Contours $C_{1},C_{2}$.\label{fig:Contours-3}}
\end{figure}

\begin{figure}
\begin{centering}
\includegraphics[scale=0.5]{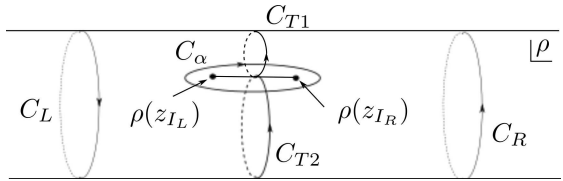}
\par\end{centering}

\protect\caption{Contours $C_{T1},C_{T2},C_{\alpha},C_{R},C_{L}$. \label{fig:Contours}}
\end{figure}

We would like to calculate the variation of $F_{N}^{(g)}(\vec{x})$
under a change of the shape of the tiny cylinder fixing the other
part of the diagram. Such a change corresponds to a change of the
moduli parameters and it induces a variation
\[
\rho(z)\to\rho(z)+\delta\rho(z)\,,
\]
 of the function $\rho(z)$ in (\ref{eq:mandelstammulti}). The change
is parametrized by the variation of the circumference of the cylinder,
i.e. $2\pi\delta\alpha$, and $\delta\mathcal{T}_{1},\delta\mathcal{T}_{2}$
defined by
\begin{eqnarray*}
\delta\mathcal{T}_{1} & = & \int_{C_{1}}dz^{\prime}\partial\delta\rho(z^{\prime})\,,\\
\delta\mathcal{T}_{2} & = & \int_{C_{2}}dz^{\prime}\partial\delta\rho(z^{\prime})\,,
\end{eqnarray*}
where the contours $C_{1},C_{2}$ are those shown in Fig. \ref{fig:Contours-3}.
$\delta\alpha$ can also be expressed as 
\[
\delta\alpha=\oint_{C_{T1}}\frac{dz}{2\pi i}\partial\delta\rho(z)=-\oint_{C_{T2}}\frac{dz}{2\pi i}\partial\delta\rho(z)\,,
\]
where the contours $C_{T1},C_{T2}$ are those depicted in Fig. \ref{fig:Contours}. 

$\delta F_{N}^{(g)}(\vec{x})$ is given by 
\begin{eqnarray}
\delta F_{N}^{(g)}(\vec{x}) & = & -\delta\mathcal{T}_{1}\oint_{C_{T1}}\frac{d\rho}{2\pi i}\left\langle T(\rho)\mathcal{O}_{N}^{(g)}\right\rangle e^{-\frac{1-Q^{2}}{2}\Gamma}-\delta\mathcal{T}_{2}\oint_{C_{T2}}\frac{d\rho}{2\pi i}\left\langle T(\rho)\mathcal{O}_{N}^{(g)}\right\rangle e^{-\frac{1-Q^{2}}{2}\Gamma}\nonumber \\
 &  & +2\pi i\delta\alpha\oint_{C_{\alpha}}\frac{d\rho}{2\pi i}\left\langle T(\rho)\mathcal{O}_{N}^{(g)}\right\rangle e^{-\frac{1-Q^{2}}{2}\Gamma}\nonumber \\
 &  & \quad+\mathrm{c.c.}\,,\label{eq:deltaF1}
\end{eqnarray}
where the contours $C_{T1},C_{T2},C_{\alpha}$ are those indicated
in Fig. \ref{fig:Contours}. By using the transformation formula
\begin{eqnarray*}
T(\rho) & = & \frac{1}{(\partial\rho(z))^{2}}\left(T(z)-(1-Q^{2})\left\{ \rho,z\right\} \right)\,,\\
\left\{ \rho,z\right\}  & = & \frac{\partial^{3}\rho}{\partial\rho}-\frac{3}{2}\left(\frac{\partial^{2}\rho}{\partial\rho}\right)^{2}\,,
\end{eqnarray*}
contour integrals of the energy-momentum tensor are expressed as 
\begin{eqnarray*}
\oint_{C}\frac{d\rho}{2\pi i}T(\rho) & = & \oint_{C}\frac{dz}{2\pi i}\frac{1}{\partial\rho(z)}\left(T(z)-(1-Q^{2})\left\{ \rho,z\right\} \right)\,,\\
\oint_{\bar{C}}\frac{d\bar{\rho}}{2\pi i}\bar{T}(\bar{\rho}) & = & \oint_{\bar{C}}\frac{d\bar{z}}{2\pi i}\frac{1}{\partial\bar{\rho}(\bar{z})}\left(\bar{T}(\bar{z})-(1-Q^{2})\left\{ \bar{\rho},\bar{z}\right\} \right)\,.
\end{eqnarray*}
Taking the local coordinates $z,\bar{z}$ convenient for calculation,
one can evaluate the right hand side of (\ref{eq:deltaF1}). 

\begin{figure}
\begin{centering}
\includegraphics[scale=0.7]{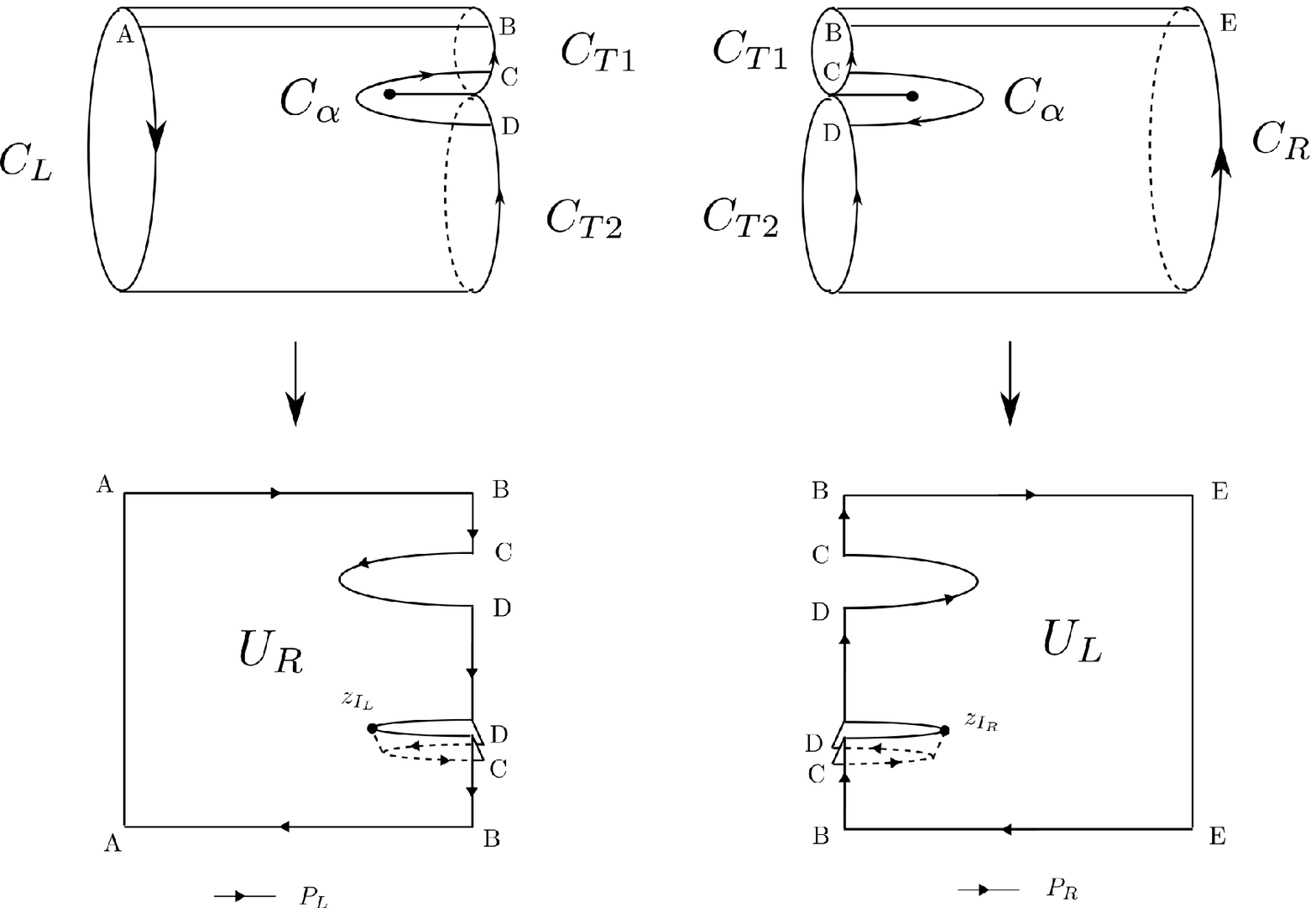}
\par\end{centering}

\protect\caption{Two pants and the paths $P_{L}\,(\mathrm{E}\to\mathrm{B}\to\mathrm{C}\to\mathrm{D}\to\mathrm{D}\to\mathrm{C}\to\mathrm{B}\to\mathrm{E})$
and $P_{R}\,(\mathrm{A}\to\mathrm{B}\to\mathrm{C}\to\mathrm{D}\to\mathrm{D}\to\mathrm{C}\to\mathrm{B}\to\mathrm{A})$.\label{fig:Two-pants-diagrams}}
\end{figure}

\begin{figure}[h]
\begin{centering}
\includegraphics[scale=0.8]{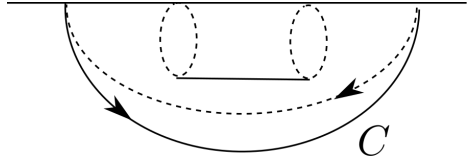}
\par\end{centering}

\protect\caption{Contour $C$.\label{fig:Contour}}
\end{figure}

Decomposing the diagram into two pants and cutting them open as in
Fig. \ref{fig:Two-pants-diagrams}, it is straightforward to show
that the right hand side of (\ref{eq:deltaF1}) is equal to 
\begin{eqnarray}
 &  & \int_{P_{L}}\frac{dz}{2\pi i}\frac{\delta\rho(z)-\delta\rho(z_{I_{L}})}{\partial\rho(z)}\left\langle \left(T(z)-(1-Q^{2})\left\{ \rho,z\right\} \right)\mathcal{O}_{N}^{(g)}\right\rangle e^{-\frac{1-Q^{2}}{2}\Gamma}\nonumber \\
 &  & +\int_{P_{R}}\frac{dz}{2\pi i}\frac{\delta\rho(z)-\delta\rho(z_{I_{R}})}{\partial\rho(z)}\left\langle \left(T(z)-(1-Q^{2})\left\{ \rho,z\right\} \right)\mathcal{O}_{N}^{(g)}\right\rangle e^{-\frac{1-Q^{2}}{2}\Gamma}\nonumber \\
 &  & \quad+\mathrm{c.c.}\,.\label{eq:intP}
\end{eqnarray}
$P_{L},P_{R}$ are the paths depicted in Fig. \ref{fig:Two-pants-diagrams}.
$\delta\rho(z)-\delta\rho(z_{I_{L}})$ and $\delta\rho(z)-\delta\rho(z_{I_{R}})$
are defined as 
\begin{eqnarray}
\delta\rho(z)-\delta\rho(z_{I_{L}}) & = & \int_{z_{I_{L}}}^{z}dz^{\prime}\partial\delta\rho(z^{\prime})\,,\nonumber \\
\delta\rho(z)-\delta\rho(z_{I_{R}}) & = & \int_{z_{I_{R}}}^{z}dz^{\prime}\partial\delta\rho(z^{\prime})\,,\label{eq:deltarho-deltarho}
\end{eqnarray}
where the paths of integration are taken to be within the regions
$U_{L}$ and $U_{R}$ in Fig. \ref{fig:Two-pants-diagrams} respectively.
Deforming the contours, we can show 
\begin{eqnarray}
\delta F_{N}^{(g)}(\vec{x}) & = & -\oint_{z_{I_{L}}}\frac{dz}{2\pi i}\frac{\delta\rho(z)-\delta\rho(z_{I_{L}})}{\partial\rho(z)}\left\langle \left(T(z)-(1-Q^{2})\left\{ \rho,z\right\} \right)\mathcal{O}_{N}^{(g)}\right\rangle e^{-\frac{1-Q^{2}}{2}\Gamma}\nonumber \\
 &  & -\oint_{z_{I_{R}}}\frac{dz}{2\pi i}\frac{\delta\rho(z)-\delta\rho(z_{I_{R}})}{\partial\rho(z)}\left\langle \left(T(z)-(1-Q^{2})\left\{ \rho,z\right\} \right)\mathcal{O}_{N}^{(g)}\right\rangle e^{-\frac{1-Q^{2}}{2}\Gamma}\nonumber \\
 &  & +\oint_{C}\frac{dz}{2\pi i}\frac{\delta\rho(z)-\delta\rho(z_{I_{L}})}{\partial\rho(z)}\left\langle \left(T(z)-(1-Q^{2})\left\{ \rho,z\right\} \right)\mathcal{O}_{N}^{(g)}\right\rangle e^{-\frac{1-Q^{2}}{2}\Gamma}\nonumber \\
 &  & -\delta\mathcal{T}_{2}\oint_{C_{R}}\frac{dz}{2\pi i}\frac{1}{\partial\rho(z)}\left\langle \left(T(z)-(1-Q^{2})\left\{ \rho,z\right\} \right)\mathcal{O}_{N}^{(g)}\right\rangle e^{-\frac{1-Q^{2}}{2}\Gamma}\nonumber \\
 &  & \quad+\mathrm{c.c.}\,,\label{eq:intC}
\end{eqnarray}
where $C$ is the contour surrounding the tiny cylinder depicted in
Fig. \ref{fig:Contour}.

In this expression, the right hand side can be evaluated once we know
the behaviors of $\rho(z),T(z)$ near the cylinder. Taking a good
local coordinate $z$ around the cylinder, the coordinate $\rho(z)$
of the light-cone diagram shall be expressed as 
\[
\rho(z)=\epsilon\tilde{\rho}(z)+\mbox{constant}\,,
\]
 and the limit to be considered is $\epsilon\to0$. In order to get
the singular behavior of $F_{N}^{(g)}(\vec{x})$ in the limit $\epsilon\to0$,
we consider the variation $\delta F_{N}^{(g)}(\vec{x})$ under $\epsilon\to\epsilon+\delta\epsilon$.
For $\epsilon\ll1$ and $z$ close to an interaction point $z_{I}$,
\begin{eqnarray*}
\delta\rho(z)-\delta\rho(z_{I}) & \sim & \frac{\delta\epsilon}{2}\partial^{2}\tilde{\rho}(z_{I})(z-z_{I})^{2}\,,\\
\partial\rho(z) & \sim & \epsilon\partial^{2}\tilde{\rho}(z_{I})(z-z_{I})\,,\\
T(z)\mathcal{O}_{N}^{(g)} & \sim & \left(\frac{\frac{3}{2}}{(z-z_{I})^{2}}+\cdots\right)\mathcal{O}_{N}^{(g)}\,,\\
\left\{ \rho,z\right\}  & \sim & \frac{-\frac{3}{2}}{(z-z_{I})^{2}}+\cdots\,,
\end{eqnarray*}
and
\begin{equation}
-\oint_{z_{I}}\frac{dz}{2\pi i}\frac{\delta\rho(z)-\delta\rho(z_{I})}{\partial\rho(z)}\left\langle \left(T(z)-(1-Q^{2})\left\{ \rho,z\right\} \right)\mathcal{O}_{N}^{(g)}\right\rangle e^{-\frac{1-Q^{2}}{2}\Gamma}\sim-\frac{3}{4}(2-Q^{2})\frac{\delta\epsilon}{\epsilon}F_{N}^{(g)}(\vec{x})\,.\label{eq:intzI}
\end{equation}
(\ref{eq:specialrho23}) implies that $\tilde{\rho}(z)$ has a simple
pole at the degenerating puncture and $C$ is a contour around it.
Using these facts, we obtain
\begin{eqnarray}
 &  & \oint_{C}\frac{dz}{2\pi i}\frac{\delta\rho(z)-\delta\rho(z_{I_{L}})}{\partial\rho(z)}\left\langle \left(T(z)-(1-Q^{2})\left\{ \rho,z\right\} \right)\mathcal{O}_{N}^{(g)}\right\rangle e^{-\frac{1-Q^{2}}{2}\Gamma}\nonumber \\
 &  & \quad\sim\frac{\delta\epsilon}{\epsilon}\oint_{C}\frac{dz}{2\pi i}z\left\langle T(z)\mathcal{O}_{N}^{(g)}\right\rangle e^{-\frac{1-Q^{2}}{2}\Gamma}\nonumber \\
 &  & \quad\sim0\,,\label{eq:collapsing}
\end{eqnarray}
 because the momentum flowing through the collapsing neck is $0$.
The fourth term on the right hand side of (\ref{eq:intC}) is of order
$\delta\epsilon$. Therefore we get
\[
\delta F_{N}^{(g)}(\vec{x})\sim\left(-6+3Q^{2}\right)\frac{\delta\epsilon}{\epsilon}F_{N}^{(g)}(\vec{x})\,,
\]
 from which we can deduce that $F_{N}^{(g)}(\vec{x})$ is expressed
as 
\begin{equation}
F_{N}^{(g)}(\vec{x})\sim\epsilon^{-6+3Q^{2}}\times\mathrm{constant}\,,\label{eq:scaling1}
\end{equation}
for $\epsilon\sim0$. 

In general, the behavior of $F_{N}^{(g)}(\vec{x})$ in the limit where
subregions of the diagram shrink to points can be studied in the same
way. The variation $\delta F_{N}^{(g)}$ under a change of the shape
of the diagram can be expressed as a sum of contour integrals of correlation
functions with energy-momentum tensor insertions. Decomposing the
diagram into pants, expressing the integrals in terms of those around
the pants and deforming the contours of the integrations, $\delta F_{N}^{(g)}(\vec{x})$
can be expressed by contour integrals in the shrinking subregions.
It is possible to evaluate them taking coordinates convenient for
describing those regions and deduce the singular behavior of $F_{N}^{(g)}(\vec{x})$
in the limit.

\subsubsection{Singular behavior of $F_{N}^{(g)}(\vec{x})$ associated with the
configuration depicted in Fig. \ref{fig:nonseparating1}\label{sub:Singular-behavior-of}}

\begin{figure}
\begin{centering}
\includegraphics[scale=0.5]{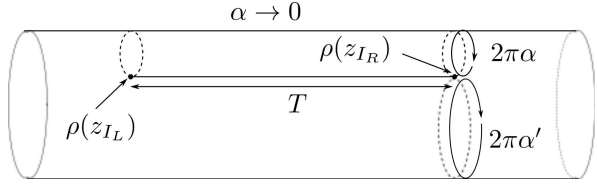}
\par\end{centering}

\protect\caption{Fig. \ref{fig:nonseparating1} embedded in a light-cone diagram. \label{fig:The-degenration-of} }
\end{figure}

As another example, let us consider the degeneration in which the
light-cone diagram develops a cylinder with vanishing width. Suppose
that the cylinder is embedded in the diagram as illustrated in Fig.
\ref{fig:The-degenration-of}. We take the limit $\alpha\to0$ with
$T$ fixed. 

\begin{figure}
\begin{centering}
\includegraphics[scale=0.5]{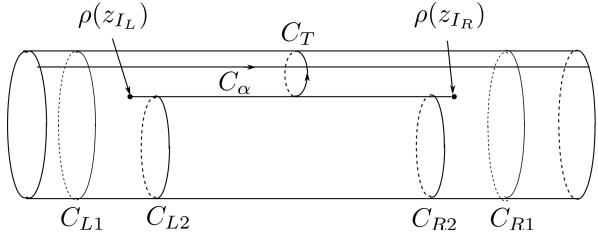}
\par\end{centering}

\protect\caption{Contours. \label{fig:Contoursthin}}
\end{figure}

In order to get the singular behavior of $F_{N}^{(g)}(\vec{x})$ in
the limit $\alpha\to0$, we evaluate the variation of $F_{N}^{(g)}$
under $\alpha\to\alpha+\delta\alpha$ which is given by
\begin{equation}
\delta F_{N}^{(g)}(\vec{x})=2\pi i\delta\alpha\oint_{C_{\alpha}}\frac{dz}{2\pi i}\frac{1}{\partial\rho(z)}\left\langle \left(T(z)-(1-Q^{2})\left\{ \rho,z\right\} \right)\mathcal{O}_{N}^{(g)}\right\rangle e^{-\frac{1-Q^{2}}{2}\Gamma}+\mathrm{c.c.}\,,\label{eq:deltaF2}
\end{equation}
where the contour $C_{\alpha}$ is shown in Fig. \ref{fig:Contoursthin}.
$C_{\alpha}$ either closes or ends in punctures. We decompose the
relevant part of the diagram into two pants which are the regions
bounded by the curves $C_{L1},C_{L2},C_{T}$ and $C_{R1},C_{R2},C_{T}$
respectively and a cylindrical region bounded by $C_{L2},C_{R2}$
in Fig. \ref{fig:Contoursthin}. We also introduce a local coordinate
$z_{L}$ around the interaction point $\rho(z_{I_{L}})$ such that
\[
\rho(z_{L})\sim\alpha(z_{L}-1-\ln z_{L})+\rho(z_{I_{L}})\,,
\]
and similarly $z_{R}$ around $\rho(z_{I_{R}})$ such that
\[
\rho(z_{R})\sim\alpha(-z_{R}+1+\ln z_{R})+\rho(z_{I_{R}})\,,
\]
for $\alpha\ll1$. We take the contour $C_{T}$ to be along the curve
\begin{eqnarray*}
\left|z_{L}\right| & \sim & \exp\left(-\frac{T_{L}}{\alpha}-1\right)\,,\\
\left|z_{R}\right| & \sim & \exp\left(-\frac{T_{R}}{\alpha}-1\right)\,,
\end{eqnarray*}
with 
\[
T_{L}+T_{R}=T\,.
\]

\begin{figure}
\begin{centering}
\includegraphics[scale=0.5]{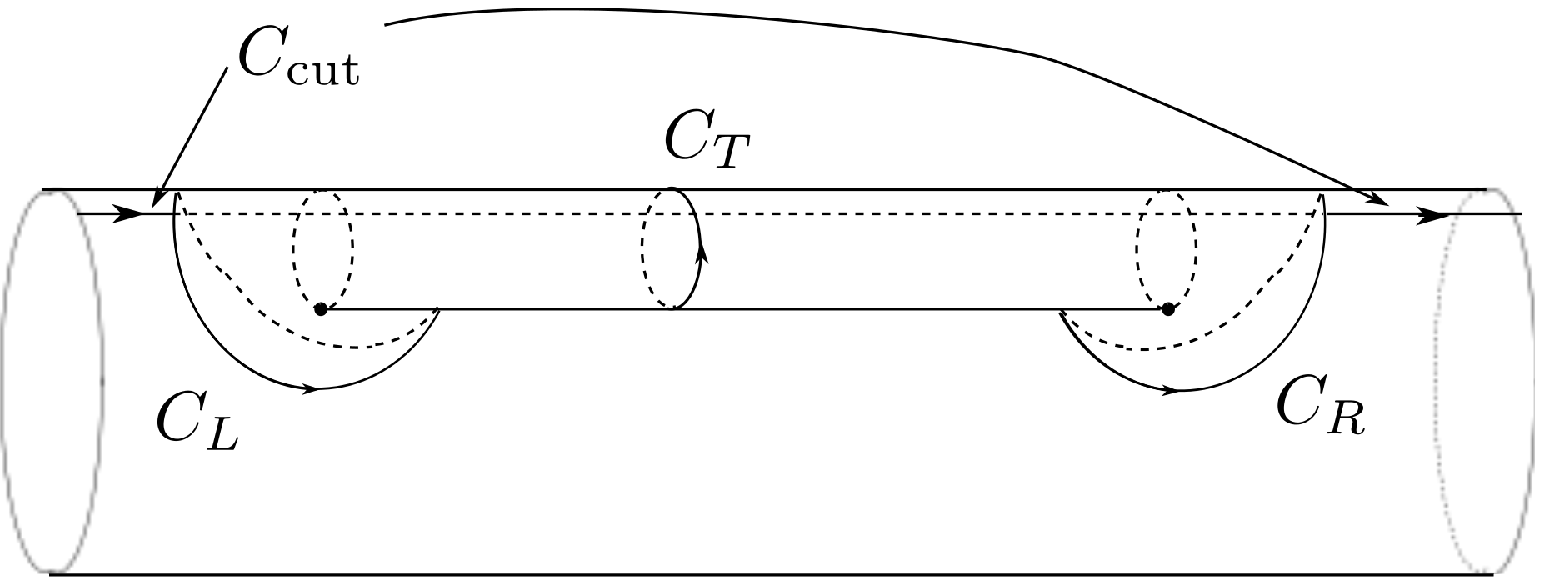}
\par\end{centering}

\protect\caption{Contours $C_{L},C_{R},C_{\mathrm{cut}}$.\label{fig:Contours-.}}
\end{figure}

Proceeding as in (\ref{eq:intP}), (\ref{eq:intC}), we get
\begin{eqnarray}
\delta F_{N}^{(g)}(\vec{x}) & = & -\oint_{1}\frac{dz_{L}}{2\pi i}\frac{\delta\rho(z_{L})-\delta\rho(z_{I_{L}})}{\partial\rho(z_{L})}\left\langle \left(T(z)-(1-Q^{2})\left\{ \rho,z\right\} \right)\mathcal{O}_{N}^{(g)}\right\rangle e^{-\frac{1-Q^{2}}{2}\Gamma}\nonumber \\
 &  & +\int_{C_{T}}\frac{dz_{L}}{2\pi i}\frac{\delta\rho(z_{L})-\delta\rho(z_{I_{L}})}{\partial\rho(z_{L})}\left\langle \left(T(z)-(1-Q^{2})\left\{ \rho,z\right\} \right)\mathcal{O}_{N}^{(g)}\right\rangle e^{-\frac{1-Q^{2}}{2}\Gamma}\nonumber \\
 &  & +\int_{C_{L}}\frac{dz_{L}}{2\pi i}\frac{\delta\rho(z_{L})-\delta\rho(z_{I_{L}})}{\partial\rho(z_{L})}\left\langle \left(T(z)-(1-Q^{2})\left\{ \rho,z\right\} \right)\mathcal{O}_{N}^{(g)}\right\rangle e^{-\frac{1-Q^{2}}{2}\Gamma}\nonumber \\
 &  & -\oint_{1}\frac{dz_{R}}{2\pi i}\frac{\delta\rho(z_{R})-\delta\rho(z_{I_{R}})}{\partial\rho(z_{R})}\left\langle \left(T(z)-(1-Q^{2})\left\{ \rho,z\right\} \right)\mathcal{O}_{N}^{(g)}\right\rangle e^{-\frac{1-Q^{2}}{2}\Gamma}\nonumber \\
 &  & -\int_{C_{T}}\frac{dz_{R}}{2\pi i}\frac{\delta\rho(z_{R})-\delta\rho(z_{I_{R}})}{\partial\rho(z_{R})}\left\langle \left(T(z)-(1-Q^{2})\left\{ \rho,z\right\} \right)\mathcal{O}_{N}^{(g)}\right\rangle e^{-\frac{1-Q^{2}}{2}\Gamma}\nonumber \\
 &  & +\int_{C_{R}}\frac{dz_{R}}{2\pi i}\frac{\delta\rho(z_{R})-\delta\rho(z_{I_{R}})}{\partial\rho(z_{R})}\left\langle \left(T(z)-(1-Q^{2})\left\{ \rho,z\right\} \right)\mathcal{O}_{N}^{(g)}\right\rangle e^{-\frac{1-Q^{2}}{2}\Gamma}\nonumber \\
 &  & +2\pi i\delta\alpha\oint_{C_{\mathrm{cut}}}\frac{dz}{2\pi i}\frac{1}{\partial\rho(z)}\left\langle \left(T(z)-(1-Q^{2})\left\{ \rho,z\right\} \right)\mathcal{O}_{N}^{(g)}\right\rangle e^{-\frac{1-Q^{2}}{2}\Gamma}\nonumber \\
 &  & \quad+\mathrm{c.c.}\,,\label{eq:Calpha1L}
\end{eqnarray}
where $C_{\mathrm{cut}}$ denotes parts of the contour $C_{\alpha}$
presented in Fig. \ref{fig:Contours-.}, together with $C_{L},C_{R}$.
The first term on the right hand side of (\ref{eq:Calpha1L}) can
be evaluated in the same way as (\ref{eq:intzI}) and we get 
\[
-\oint_{1}\frac{dz_{L}}{2\pi i}\frac{\delta\rho(z_{L})-\delta\rho(z_{I_{L}})}{\partial\rho(z_{L})}\left\langle \left(T(z)-(1-Q^{2})\left\{ \rho,z\right\} \right)\mathcal{O}_{N}^{(g)}\right\rangle e^{-\frac{1-Q^{2}}{2}\Gamma}\sim-\frac{3}{4}(2-Q^{2})\frac{\delta\alpha}{\alpha}F_{N}^{(g)}\,.
\]
The second term is evaluated to be
\begin{eqnarray}
 &  & \int_{C_{T}}\frac{dz_{L}}{2\pi i}\frac{\delta\rho(z_{L})-\delta\rho(z_{I_{L}})}{\partial\rho(z_{L})}\left\langle \left(T(z)-(1-Q^{2})\left\{ \rho,z\right\} \right)\mathcal{O}_{N}^{(g)}\right\rangle e^{-\frac{1-Q^{2}}{2}\Gamma}\nonumber \\
 &  & \quad\sim-\int_{0}^{2\pi}\frac{d\sigma}{2\pi}\frac{\delta\alpha}{-\alpha}z_{L}^{^{2}}(\frac{T_{L}}{\alpha}-i\sigma)\frac{1}{z_{L}^{2}}\left\langle \left(\frac{1}{2}\left|\vec{p}\right|^{2}+Qp^{1}+\frac{1}{2}Q^{2}\right)\mathcal{O}_{N}^{(g)}\right\rangle e^{-\frac{1-Q^{2}}{2}\Gamma}\nonumber \\
 &  & \quad\sim\frac{D}{4}\frac{\delta\alpha}{\alpha}+\mbox{imaginary part}\,,\label{eq:CT}
\end{eqnarray}
where we have used the fact that the states propagating through $C_{T}$
are projected to be GSO even and the dominant contributions in the
limit $\alpha\to0$ come from the states with the momentum $p^{i}=-Q\delta_{i,1}$.
$D$ denotes the number of noncompact bosons in the world sheet theory
and $\frac{D}{4}\frac{\delta\alpha}{\alpha}$ in the last line originates
from the momentum integral. For $\alpha\ll1$, taking the contour
$C_{L}$ to be along the curve $\left|z_{L}\right|=\alpha^{-\gamma}\,(0<\gamma<1)$,
the third term on the right hand side of (\ref{eq:Calpha1L}) is evaluated
as
\begin{eqnarray}
 &  & \int_{C_{L}}\frac{dz_{L}}{2\pi i}\frac{\delta\rho(z_{L})-\delta\rho(z_{I_{L}})}{\partial\rho(z_{L})}\left\langle \left(T(z)-(1-Q^{2})\left\{ \rho,z\right\} \right)\mathcal{O}_{N}^{(g)}\right\rangle e^{-\frac{1-Q^{2}}{2}\Gamma}\nonumber \\
 &  & \quad\sim\frac{\delta\alpha}{\alpha}\oint_{C_{L}}\frac{dz_{L}}{2\pi i}z_{L}\left\langle T(z)\mathcal{O}_{N}^{(g)}\right\rangle e^{-\frac{1-Q^{2}}{2}\Gamma}+\mathrm{constant}\times\delta(\alpha^{\gamma})\nonumber \\
 &  & \quad\sim-\frac{1}{2}Q^{2}\frac{\delta\alpha}{\alpha}F_{N}^{(g)}\,,\label{eq:jump}
\end{eqnarray}
where we have ignored the term of the order $\delta(\alpha^{\gamma})$.
The states propagating through $C_{L}$ are GSO odd and the dominant
contributions to the contour integral come from the states with the
momentum $p^{i}=-Q\delta_{i,1}$. The fourth to the sixth terms are
evaluated in the same way, taking the contour $C_{R}$ to be along
the curve $\left|z_{R}\right|=\alpha^{-\gamma}$. The singular contributions
of the integration along $C_{\mathrm{cut}}$ can come from the regions
near the contours $C_{R},C_{L}$ and
\begin{eqnarray}
 &  & 2\pi i\delta\alpha\oint_{C_{\mathrm{cut}}}\frac{dz}{2\pi i}\frac{1}{\partial\rho(z)}\left\langle \left(T(z)-(1-Q^{2})\left\{ \rho,z\right\} \right)\mathcal{O}_{N}^{(g)}\right\rangle e^{-\frac{1-Q^{2}}{2}\Gamma}\nonumber \\
 &  & \quad\sim2\pi i\delta\alpha\int^{\alpha^{-\gamma}}\frac{dz_{L}}{2\pi i}\frac{1}{\partial\rho(z_{L})}\left\langle \left(T(z)-(1-Q^{2})\left\{ \rho,z\right\} \right)\mathcal{O}_{N}^{(g)}\right\rangle e^{-\frac{1-Q^{2}}{2}\Gamma}\nonumber \\
 &  & \hphantom{\quad\sim}+2\pi i\delta\alpha\int_{\alpha^{-\gamma}}\frac{dz_{R}}{2\pi i}\frac{1}{\partial\rho(z_{R})}\left\langle \left(T(z)-(1-Q^{2})\left\{ \rho,z\right\} \right)\mathcal{O}_{N}^{(g)}\right\rangle e^{-\frac{1-Q^{2}}{2}\Gamma}\nonumber \\
 &  & \quad\sim\mathrm{constant}\times\delta(\alpha^{2\gamma})\,,\label{eq:cut}
\end{eqnarray}
which can be ignored in the limit $\alpha\to0$. Putting these altogether,
the right hand side is evaluated to be 
\[
\left(-6+Q^{2}+\frac{D}{2}\right)\frac{\delta\alpha}{\alpha}F_{N}^{(g)}(\vec{x})\,.
\]
From this, we can deduce that $F_{N}^{(g)}(\vec{x})$ is expressed
as 
\begin{equation}
F_{N}^{(g)}(\vec{x})\sim\alpha^{-6+Q^{2}+\frac{D}{2}}\times\mbox{constant}\,,\label{eq:scaling2}
\end{equation}
for $\alpha\ll1$.

\subsubsection{Collisions of interaction points}

\begin{figure}
\begin{centering}
\includegraphics[scale=0.5]{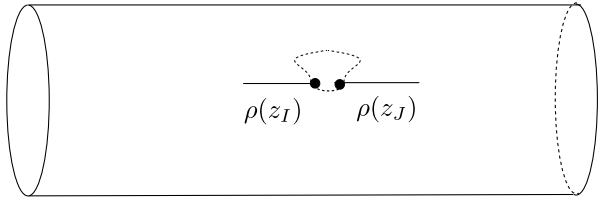}
\par\end{centering}

\protect\caption{\label{fig:scaling4}Two interaction points come close to each other
without degeneration. The coordinate size of the neck does not go
to zero in the limit $\rho(z_{I})\to\rho(z_{J})$. }
\end{figure}

The technique developed above is applicable to the situation in which
the interaction points come close to each other but no degeneration
occurs. When two of the interaction points come close to each other
as shown in Fig. \ref{fig:scaling4}, it is possible to take a local
coordinate $z$ around the interaction points so that $\rho(z)$ can
be expressed as 
\begin{equation}
\rho(z)\sim\epsilon(z^{3}-3z)+\mathrm{constant}\,,\label{eq:rho4}
\end{equation}
where the limit we should consider is $\epsilon\to0$. 

\begin{figure}
\begin{centering}
\includegraphics[scale=0.5]{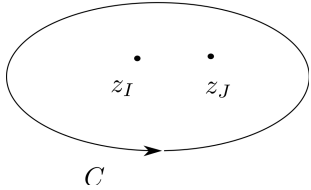}
\par\end{centering}

\protect\caption{\label{fig:zIzJC}$z_{I},z_{J},C$.}
\end{figure}

The variation of $F_{N}^{(g)}(\vec{x})$ under $\epsilon\to\epsilon+\delta\epsilon$
can be given as
\begin{eqnarray}
\delta F_{N}^{(g)}(\vec{x}) & \sim & -\oint_{z_{I}}\frac{dz}{2\pi i}\frac{\delta\rho(z)-\delta\rho(z_{I})}{\partial\rho(z)}\left\langle \left(T(z)-(1-Q^{2})\left\{ \rho,z\right\} \right)\mathcal{O}_{N}^{(g)}\right\rangle e^{-\frac{1-Q^{2}}{2}\Gamma}\nonumber \\
 &  & -\oint_{z_{J}}\frac{dz}{2\pi i}\frac{\delta\rho(z)-\delta\rho(z_{J})}{\partial\rho(z)}\left\langle \left(T(z)-(1-Q^{2})\left\{ \rho,z\right\} \right)\mathcal{O}_{N}^{(g)}\right\rangle e^{-\frac{1-Q^{2}}{2}\Gamma}\nonumber \\
 &  & +\oint_{C}\frac{dz}{2\pi i}\frac{\delta\rho(z)}{\partial\rho(z)}\left\langle \left(T(z)-(1-Q^{2})\left\{ \rho,z\right\} \right)\mathcal{O}_{N}^{(g)}\right\rangle e^{-\frac{1-Q^{2}}{2}\Gamma}\nonumber \\
 &  & \quad+\mbox{c.c.}\,,\label{eq:deltaF3}
\end{eqnarray}
where $z_{I},z_{J},C$ are depicted in Fig. \ref{fig:zIzJC}. The
terms in the first and the second lines can be evaluated as in (\ref{eq:intzI})
and we obtain $\sim-\frac{3}{4}(2-Q^{2})\frac{\delta\epsilon}{\epsilon}F_{N}^{(g)}$.
With (\ref{eq:rho4}), we get 
\begin{eqnarray*}
\left\{ \rho,z\right\}  & \sim & \frac{-4}{z^{2}}\,,\\
\frac{\delta\rho(z)}{\partial\rho(z)} & \sim & \frac{\delta\epsilon}{3\epsilon}z\,,
\end{eqnarray*}
for $z\gg1$ and the term in the third line is evaluated to be $\sim\frac{4}{3}(1-Q^{2})\frac{\delta\epsilon}{\epsilon}F_{N}^{(g)}$.
Therefore we eventually get 
\begin{equation}
F_{N}^{(g)}(\vec{x})\sim\epsilon^{-\frac{10}{3}+\frac{1}{3}Q^{2}}\times\mbox{constant}\,.\label{eq:scaling3}
\end{equation}

The case in which $n$ interaction points come close to each other
can be treated in the same way. With a good local coordinate $z$,
$\rho(z)$ can be expressed as
\[
\rho(z)\sim\epsilon(z^{n+1}+\cdots)+\mathrm{constant}\,,
\]
and we get
\begin{equation}
F_{N}^{(g)}(\vec{x})\sim\epsilon^{\frac{1}{n+1}(-2n^{2}-n+\frac{1}{2}(n^{2}-n)Q^{2})}\times\mathrm{constant}\,.\label{eq:scaling4}
\end{equation}

\subsection{Divergences of the amplitudes}

Using eqs.(\ref{eq:scaling1}), (\ref{eq:scaling2}), (\ref{eq:scaling3}),
(\ref{eq:scaling4}), we can check if the integrations around the
singularities studied in the previous subsections give divergent contributions
to the amplitude $A_{N}^{(g)}$. For example, in the case of the configuration
presented in Fig. \ref{fig:scaling1}, the relevant part of the integration
measure is expressed as
\[
\int d\epsilon\epsilon^{2}\,,
\]
for $\epsilon\ll1$, and with (\ref{eq:scaling1}) the contribution
to the amplitude from the neighborhood of the singularity goes as
\[
\int d\epsilon\epsilon^{-4+3Q^{2}}\,.
\]
This integral diverges when $Q=0$ but converges if $Q^{2}$ is large
enough. The same happens for other configurations discussed in the
previous subsection. Notice that the infinitely thin cylinder in Fig.
\ref{fig:The-degenration-of} leads to a divergence in spite of the
Feynman $i\varepsilon$, because of the contributions from the tiny
regions at the ends.

Therefore the amplitudes diverge in the superstring field theory with
$Q=0$, which is the theory in the critical dimension. The Feynman
$i\varepsilon$ is not enough to make them finite, partly because
some of the configurations correspond to the ultraviolet region with
respect to the world sheet metric $ds^{2}=d\rho d\bar{\rho}$. The
divergences are also due to the presence of $T_{F}$ and $\bar{T}_{F}$
at the interaction points. In order to make sense out of the string
field theory, we need to regularize the divergences. As we have seen
in the examples discussed here, it seems that we can do so by taking
$Q^{2}$ large enough.

\section{Regularization of divergences \label{sec:Regularization-of-the}}

In this section, we would like to show that by taking $\varepsilon>0$
and $Q^{2}>10$ the amplitude (\ref{eq:ANlin}) becomes finite. The
singularities coming from cylinders with infinite length and nonvanishing
width are taken care of by taking $\varepsilon>0$. Other types of
singularities correspond to light-cone diagrams which involve infinitely
thin cylinders and/or colliding interaction points. As we have seen
in the previous section, the singularities of $F_{N}^{(g)}$ can be
deduced from the behavior of $\rho(z),\,\bar{\rho}(\bar{z})$ in tiny
regions around the relevant interaction points. The singular configuration
corresponds to the limit where these regions shrink to points. 

\begin{figure}
\begin{centering}
\includegraphics[scale=0.5]{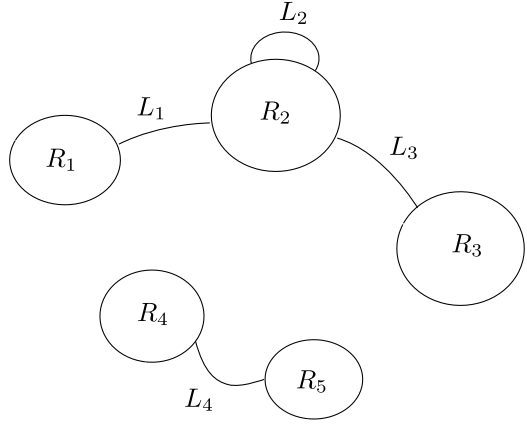}
\par\end{centering}

\protect\caption{Regions shrinking to points connected by infinitely thin tubes.\label{fig:Regions-shrinking-to}}
\end{figure}

General singular configurations we should deal with can be realized
in the following way:
\begin{itemize}
\item Let $G$ be a subregion of a regular light-cone diagram which consists
of regions $R_{a}\,(a=1,2,\cdots)$ connected by propagators $L_{b}\,(b=1,2,,\cdots)$. 
\item The singular configuration corresponds to the limit in which the regions
$R_{a}$ shrink to points and the cylinders $L_{b}$ become infinitely
thin, as illustrated in Fig. \ref{fig:Regions-shrinking-to}.
\end{itemize}
In order to study the singular behavior of $F_{N}^{(g)}(\vec{x})$
in such a limit, it is convenient to take the integration variables
$\vec{x}=(x^{1},x^{2},\cdots,x^{n})\in\mathbb{R}^{n}$ in the following
way. Let $x^{1},\cdots,x^{n_{G}}$ be the independent linear combinations
of differences of coordinates $\rho,\bar{\rho}$ of the interaction
points and magnitudes of jump discontinuities of $\rho,\bar{\rho}$
in the regions $R_{a}$ so that the limit where they shrink to points
and the cylinders $L_{b}$ become infinitely thin is represented by
$x^{j}\to0\,(j=1,\cdots,n_{G})$. We take $x^{n_{G}+1},\cdots,x^{n}$
to represent the shape of the light-cone diagram outside of $G$ and
the positions of $R_{a}$. 

Then the singularity in question is at 
\[
\vec{x}=(\overbrace{0,\cdots,0}^{n_{G}\mbox{ times}},\vec{y})\,,
\]
where
\[
\vec{y}\equiv(y^{n_{G}+1},\cdots,y^{n})\,.
\]
In order to study the behavior of $F_{N}^{(g)}(\vec{x})$ at the singularity,
we estimate 
\[
F_{N}^{(g)}\left(\epsilon\vec{w},\vec{y}\right)
\]
in the limit $\epsilon\to0$ with
\begin{equation}
\vec{w}=(w^{1},\cdots,w^{n_{G}})\ (\left|\vec{w}\right|=\sqrt{(w^{1})^{2}+(w^{2})^{2}+\cdots+(w^{n_{G}})^{2}}=1)\,,\label{eq:vecw}
\end{equation}
fixed. We can do so as was done in the previous section, if $(\epsilon\vec{w},\vec{y})$
itself does not correspond to a singular configuration, i.e. 
\[
\vec{v}_{k}\cdot(\epsilon\vec{w},\vec{y})\ne0
\]
for any $k$. This happens for a generic choice of $\vec{w}$. We
would like to first analyze $F_{N}^{(g)}\left(\epsilon\vec{w},\vec{y}\right)$
for such $\vec{w}$ in the limit $\epsilon\to0$. 

\subsection{$F_{N}^{(g)}\left(\epsilon\vec{w},\vec{y}\right)$ in the limit $\epsilon\to0$}

For $\epsilon\ll1$, with a good local coordinate $z$ on $R_{a}$,
the $\rho(z)$ is approximated as
\[
\rho(z)\sim\epsilon\rho_{a}(z)+\mbox{constant}\,,
\]
where $\rho_{a}(z)$ is a multivalued meromorphic function on $R_{a}$.
Suppose that the region $R_{a}$ has genus $g_{a}$ and $k_{a}+l_{a}$
boundaries. $l_{a}$ of the boundaries are associated with the thin
cylinders attached to $R_{a}$ and $k_{a}$ of them are around the
necks which connect $R_{a}$ with the rest of the surface. $l_{a}$
boundaries associated with the thin cylinders correspond to simple
poles of $\partial\rho_{a}(z)$ and other $k_{a}$ boundaries correspond
to higher order poles at $z=z_{i}^{(a)}\,(i=1,\cdots,k_{a})$. We
assume that for $z\sim z_{i}^{(a)}$, $\partial\rho_{a}(z)$ behaves
as

\[
\partial\rho_{a}(z)\sim\frac{r_{i}^{(a)}}{(z-z_{i}^{(a)})^{n_{i}^{(a)}}}+\cdots\,.
\]
$(n_{i}^{(a)}\geq2)$. Since the degree of the differential $\partial\rho_{a}(z)dz$
should be $2g_{a}-2$, we get 
\begin{equation}
N_{I}^{(a)}-\sum_{i=1}^{k_{a}}n_{i}^{(a)}-l_{a}=2g_{a}-2\,,\label{eq:degreederrho}
\end{equation}
where $N_{I}^{(a)}$ is the number of the interaction points included
in $R_{a}$. In order for the statement that $R_{a}$ shrinks to a
point to make sense, $N_{I}^{(a)}\geq1$.

Now let us calculate the behavior of $F_{N}^{(g)}\left(\epsilon\vec{w},\vec{y}\right)$
in the limit $\epsilon\to0$. The variation of $F_{N}^{(g)}\left(\epsilon\vec{w},\vec{y}\right)$
under $\epsilon\to\epsilon+\delta\epsilon$ can be evaluated as in
the examples discussed in the previous section. Expressing the variation
$\delta F_{N}^{(g)}$ as a sum of contour integrals of correlation
functions with energy-momentum tensor insertions and deforming the
contours, we eventually get 
\begin{eqnarray}
\delta F_{N}^{(g)}\left(\epsilon\vec{w},\vec{y}\right) & \sim & -\sum_{I_{G}}\oint_{z_{I_{G}}}\frac{dz}{2\pi i}\frac{\delta\rho(z)-\delta\rho(z_{I_{G}})}{\partial\rho(z)}\left\langle \left(T(z)-(1-Q^{2})\left\{ \rho,z\right\} \right)\mathcal{O}_{N}^{(g)}\right\rangle e^{-\frac{1-Q^{2}}{2}\Gamma}\nonumber \\
 &  & +\sum_{a}\sum_{b}\int_{C_{b}^{(a)}}\frac{dz}{2\pi i}\frac{\delta\rho(z)}{\partial\rho(z)}\left\langle \left(T(z)-(1-Q^{2})\left\{ \rho,z\right\} \right)\mathcal{O}_{N}^{(g)}\right\rangle e^{-\frac{1-Q^{2}}{2}\Gamma}\nonumber \\
 &  & +\sum_{a}\sum_{i}\int_{C_{i}^{(a)}}\frac{dz}{2\pi i}\frac{\delta\rho(z)}{\partial\rho(z)}\left\langle \left(T(z)-(1-Q^{2})\left\{ \rho,z\right\} \right)\mathcal{O}_{N}^{(g)}\right\rangle e^{-\frac{1-Q^{2}}{2}\Gamma}\nonumber \\
 &  & +\sum_{C_{\mathrm{cut}}}\int_{C_{\mathrm{cut}}}\frac{dz}{2\pi i}\frac{\Delta\delta\rho}{\partial\rho(z)}\left\langle \left(T(z)-(1-Q^{2})\left\{ \rho,z\right\} \right)\mathcal{O}_{N}^{(g)}\right\rangle e^{-\frac{1-Q^{2}}{2}\Gamma}\nonumber \\
 &  & \quad+\mathrm{c.c.}\,.\label{eq:deltageneral}
\end{eqnarray}
Here $z=z_{I_{G}}\,(I_{G}=1,2,\cdots)$ correspond to the interaction
points included in $G$, the contour $C_{b}^{(a)}$ denotes the one
along the boundary of $R_{a}$ around $L_{b}$ and $C_{i}^{(a)}$
denotes a contour along the $i$-th boundary of $R_{a}$ which corresponds
to the pole $z_{i}^{(a)}$ of $\partial\rho_{a}$. The terms in the
fourth line come from the possible multivaluedness of $\delta\rho$
and we take $\delta\rho$ to have a jump $\Delta\delta\rho$ along
the contour $C_{\mathrm{cut}}$. There can be contributions from integrations
along contours outside of $G$, but they correspond to the terms in
$F_{N}^{(g)}$ which vanish in the limit $\epsilon\to0$.

The right hand side of (\ref{eq:deltageneral}) can be evaluated as
was done in the previous section. Each term in the first line can
be evaluated as in (\ref{eq:intzI}): 
\[
-\oint_{z_{I}}\frac{dz}{2\pi i}\frac{\delta\rho(z)-\delta\rho(z_{I})}{\partial\rho(z)}\left\langle \left(T(z)-(1-Q^{2})\left\{ \rho,z\right\} \right)\mathcal{O}_{N}^{(g)}\right\rangle e^{-\frac{1-Q^{2}}{2}\Gamma}\sim-\frac{3}{4}(2-Q^{2})\frac{\delta\epsilon}{\epsilon}F_{N}^{(g)}\,.
\]
The terms in the second line can be estimated as in (\ref{eq:CT}):
\begin{eqnarray*}
 &  & \sum_{a}\int_{C_{b}^{(a)}}\frac{dz}{2\pi i}\frac{\delta\rho(z)}{\partial\rho(z)}\left\langle \left(T(z)-(1-Q^{2})\left\{ \rho,z\right\} \right)\mathcal{O}_{N}^{(g)}\right\rangle e^{-\frac{1-Q^{2}}{2}\Gamma}\\
 &  & \quad\quad\sim\frac{D}{4}\frac{\delta\alpha}{\alpha}+\mbox{imaginary part}\,.
\end{eqnarray*}
 The terms in the third line can be calculated by using 
\begin{eqnarray}
\frac{\delta\rho_{a}(z)}{\partial\rho_{a}(z)} & \sim & -\frac{\delta\epsilon}{\epsilon}\frac{z-z_{i}^{(a)}}{n_{i}^{(a)}-1}\,,\nonumber \\
\left\{ \rho,z\right\}  & \sim & -\frac{1}{2}\frac{(n_{i}^{(a)}-1)^{2}-1}{(z-z_{i}^{(a)})^{2}}\,,\nonumber \\
\left\langle T(z)\mathcal{O}_{N}^{(g)}\right\rangle e^{-\frac{1-Q^{2}}{2}\Gamma} & \sim & \frac{1}{(z-z_{i}^{(a)})^{2}}\left\langle \left(\frac{1}{2}\left|\vec{p}\right|^{2}+Qp^{1}+\cdots\right)\mathcal{O}_{N}^{(g)}\right\rangle e^{-\frac{1-Q^{2}}{2}\Gamma}\,,\label{eq:zIa}
\end{eqnarray}
for $z\sim z_{i}^{(a)}$. The factor which appears on the right hand
side of (\ref{eq:zIa}) can be estimated by using
\[
\frac{1}{2}\left|\vec{p}\right|^{2}+Qp^{1}\geq\begin{cases}
0 & k_{a}+l_{a}=1\\
-\frac{Q^{2}}{2} & k_{a}+l_{a}\geq2
\end{cases}\,.
\]
As in the examples in the previous section, the terms in the fourth
line of (\ref{eq:deltageneral}) can give only negligible contributions
to $F_{N}^{(g)}$. From these, we can see that for $\epsilon\ll1$,
$F_{N}^{(g)}\left(\epsilon\vec{w},\vec{y}\right)$ behaves as 
\begin{equation}
F_{N}^{(g)}\left(\epsilon\vec{w},\vec{y}\right)\sim\epsilon^{\gamma_{G}}\times\mbox{constant}\,,\label{eq:partition}
\end{equation}
where
\begin{eqnarray}
\gamma_{G} & = & \sum_{a}\gamma_{R_{a}}\,,\nonumber \\
\gamma_{R_{a}} & \geq & -3N_{I}^{(a)}+\sum_{i=1}^{k_{a}}\left(n_{i}^{(a)}-1-\frac{1}{n_{i}^{(a)}-1}\right)+Q^{2}\left[\frac{3}{2}N_{I}^{(a)}-\sum_{i=1}^{k_{a}}\left(n_{i}^{(a)}-1-\frac{\delta_{k_{a}+l_{a},1}}{n_{i}^{(a)}-1}\right)\right]\,,\label{eq:gammaRa}
\end{eqnarray}
if $(\epsilon\vec{w},\vec{y})$ does not correspond to a singular
configuration.

\subsection{Proof of finiteness of $A_{N}^{(g)}$}

If $\gamma_{G}<0$, $F_{N}^{(g)}\left(\vec{x}\right)$ is singular
at $\vec{x}=(0,\vec{y})$. Let us show that we can make $\gamma_{R_{a}}>0$
by choosing $Q^{2}$ large enough. 

When $k_{a}+l_{a}\geq2$, we have
\begin{equation}
\gamma_{R_{a}}\geq-3N_{I}^{(a)}+\sum_{i=1}^{k_{a}}\left(n_{i}^{(a)}-1-\frac{1}{n_{i}^{(a)}-1}\right)+Q^{2}\left[\frac{3}{2}N_{I}^{(a)}-\sum_{i=1}^{k_{a}}\left(n_{i}^{(a)}-1\right)\right]\,.\label{eq:geq2}
\end{equation}
From (\ref{eq:degreederrho}), we get
\begin{equation}
\frac{3}{2}N_{I}^{(a)}-\sum_{i=1}^{k_{a}}\left(n_{i}^{(a)}-1\right)=\frac{1}{2}N_{I}^{(a)}+k_{a}+l_{a}+2g_{a}-2>0\,.\label{eq:geq2positive}
\end{equation}
 Substituting $Q^{2}=6$ into (\ref{eq:geq2}) yields
\begin{eqnarray*}
\gamma_{R_{a}} & \geq & 6N_{I}^{(a)}-5\sum_{i=1}^{k_{a}}\left(n_{i}^{(a)}-1\right)-\sum_{i=1}^{k_{a}}\frac{1}{n_{i}^{(a)}-1}\\
 & = & 5\left(N_{I}^{(a)}-\sum_{i=1}^{k_{a}}n_{i}^{(a)}+k_{a}\right)+N_{I}^{(a)}-\sum_{i=1}^{k_{a}}\frac{1}{n_{i}^{(a)}-1}\\
 & \geq & 5\left(k_{a}+l_{a}+2g_{a}-2\right)+N_{I}^{(a)}-k_{a}\,.
\end{eqnarray*}
Since (\ref{eq:degreederrho}) implies
\[
N_{I}^{(a)}=\sum_{i=1}^{k_{a}}n_{i}^{(a)}+l_{a}+2g_{a}-2\geq2k_{a}+l_{a}+2g_{a}-2\geq k_{a}+2g_{a}\geq k_{a}\,,
\]
we obtain $\gamma_{R_{a}}\geq0$ for $Q^{2}=6$. From (\ref{eq:geq2positive})
we can see that $\gamma_{R_{a}}>0$ holds if $Q^{2}>6$. 

When $k_{a}+l_{a}=1$, the only possibility is $k_{a}=1,\, l_{a}=0$.
(\ref{eq:gammaRa}) becomes
\[
\gamma_{R_{a}}\geq-3N_{I}^{(a)}+n_{1}^{(a)}-1-\frac{1}{n_{1}^{(a)}-1}+Q^{2}\left(\frac{3}{2}N_{I}^{(a)}-n_{1}^{(a)}+1+\frac{1}{n_{1}^{(a)}-1}\right)\,.
\]
If $g_{a}\geq1$, we can prove
\[
\frac{3}{2}N_{I}^{(a)}-n_{1}^{(a)}+1+\frac{1}{n_{1}^{(a)}-1}\geq\frac{1}{2}N_{I}^{(a)}+2g_{a}-1>0\,,
\]
 and for $Q^{2}=6$
\begin{eqnarray*}
\gamma_{R_{a}} & \geq & 6N_{I}^{(a)}-5\left(n_{1}^{(a)}-1-\frac{1}{n_{1}^{(a)}-1}\right)\\
 & > & 6N_{I}^{(a)}-5\left(n_{1}^{(a)}-1\right)\\
 & = & 5\left(2g_{a}-1\right)+N_{I}^{(a)}\\
 & > & 0\,.
\end{eqnarray*}
Therefore $\gamma_{R_{a}}>0$ for $Q^{2}>6$. If $g_{a}=0$, (\ref{eq:degreederrho})
becomes 
\[
N_{I}^{(a)}=n_{1}^{(a)}-2\,,
\]
and $\gamma_{R_{a}}$ is given by
\[
\gamma_{R_{a}}\geq\frac{1}{n_{1}^{(a)}-1}\left[-\left(2n_{1}^{(a)}-3\right)\left(n_{1}^{(a)}-2\right)+\frac{1}{2}Q^{2}\left(n_{1}^{(a)}-2\right)\left(n_{1}^{(a)}-3\right)\right]\,.
\]
In this case, in order for the statement that $R_{a}$ shrinks to
a point to make sense, $N_{I}^{(a)}\geq2$. Since $n_{1}^{(a)}=N_{I}^{(a)}+2\geq4$,
we get
\[
\left(n_{1}^{(a)}-2\right)\left(n_{1}^{(a)}-3\right)>0\,.
\]
For $Q^{2}=10$, we obtain
\[
\gamma_{R_{a}}\geq\frac{3}{n_{1}^{(a)}-1}\left(n_{1}^{(a)}-2\right)\left(n_{1}^{(a)}-4\right)\geq0\,.
\]
 Therefore $\gamma_{R_{a}}>0$ for $Q^{2}>10$. 

Thus we have proven that $\gamma_{G}=\sum\gamma_{R_{a}}>0$ holds
for any $G$, if we take $Q^{2}>10$. Since
\[
F_{N}^{(g)}\left(\epsilon\vec{w},\vec{y}\right)\sim\epsilon^{\gamma_{G}}\times\mbox{constant}\,,
\]
for generic $\vec{w}$, the fact that $\gamma_{G}>0$ for any $G$
seems to suggest that $F_{N}^{(g)}(\vec{x})$ does not have any singularities
and the amplitude is finite. We would like to prove that this is the
case in the following. 

Let us first prove that putting $F_{N}^{(g)}(\vec{0},\vec{y})=0$,
$F_{N}^{(g)}(\vec{x})$ becomes continuous at $\vec{x}=(\vec{0},\vec{y})$,
if $Q^{2}>10$. For generic $\vec{w}$, $F_{N}^{(g)}\left(\epsilon\vec{w},\vec{y}\right)$
behaves in the limit $\epsilon\to0$ as 
\[
F_{N}^{(g)}\left(\epsilon\vec{w},\vec{y}\right)\sim\epsilon^{\gamma_{G}}\times\mbox{constant}\,,
\]
with $\gamma_{G}>0$. Hence as a function of $\tilde{\epsilon}=\epsilon^{\gamma_{G}}$,
$F_{N}^{(g)}\left(\epsilon\vec{w},\vec{y}\right)$ is differentiable
at $\tilde{\epsilon}=0$. It is smooth with respect to $\vec{w},\vec{y}$
when $\tilde{\epsilon}\ne0$. Therefore we can find a constant $M>0$
such that 
\begin{equation}
\left|F_{N}^{(g)}\left(\epsilon(\vec{w}+\delta\vec{w}),\vec{y}+\delta\vec{y}\right)\right|<\epsilon^{\gamma_{G}}M\,,\label{eq:inequal}
\end{equation}
for any $\delta\vec{w},\,\delta\vec{y}$ with $\left|\delta\vec{w}\right|,\left|\delta\vec{y}\right|,\epsilon$
sufficiently small. If this holds for any $\vec{w}$, $F_{N}^{(g)}(\vec{x})$
is continuous at $\vec{x}=(\vec{0},\vec{y})$. Therefore we need to
study the case where $\vec{w}$ is not generic in the sense that $(\epsilon\vec{w},\vec{y})$
corresponds to a singular configuration, in order to prove the continuity
of $F_{N}^{(g)}(\vec{x})$

Suppose that $(\epsilon\vec{w}_{0},\vec{y})$ corresponds to a singular
configuration. It should correspond to the limit in which a subregion
$G^{\prime}$ of $G$ shrinks to a point. With a rearrangement of
the integration variables, $\vec{w}_{0}$ can be expressed as 
\[
\vec{w}_{0}=(\overbrace{0,\cdots,0}^{n_{G^{\prime}}\mbox{ times}},\vec{y}^{\prime})\,.
\]
The value of $F_{N}^{(g)}\left(\vec{x}\right)$ in the neighborhood
of the point $\vec{x}=(\epsilon\vec{w}_{0},\vec{y})$ can be studied
by estimating 
\begin{equation}
F_{N}^{(g)}\left(\epsilon_{0}(\vec{w}_{0}+\epsilon^{\prime}\vec{w}^{\prime}),\vec{y}\right)\,,\label{eq:FNg1}
\end{equation}
where
\begin{eqnarray*}
 &  & \vec{w}^{\prime}=(w^{\prime1},\cdots,w^{\prime n_{G^{\prime}}},0\cdots0)\,,\\
 &  & \left|\vec{w}^{\prime}\right|=1\,,\\
 &  & \left|\epsilon_{0}(\vec{w}_{0}+\epsilon^{\prime}\vec{w}^{\prime})\right|=\epsilon\,,
\end{eqnarray*}
and $0<\epsilon^{\prime}\ll1$. If $\vec{x}=(\epsilon_{0}(\vec{w}_{0}+\epsilon^{\prime}\vec{w}^{\prime}),\vec{y})$
corresponds to a light-cone diagram without any degenerations or collisions
of interaction points, it is straightforward to estimate (\ref{eq:FNg1})
by computing the variation $\delta F_{N}^{(g)}$ under $\epsilon^{\prime}\to\epsilon^{\prime}+\delta\epsilon^{\prime}$
using the techniques presented in the previous section and we obtain
\begin{equation}
F_{N}^{(g)}\left(\epsilon_{0}(\vec{w}_{0}+\epsilon^{\prime}\vec{w}^{\prime}),\vec{y}\right)\sim\epsilon_{0}^{\gamma_{G}}\left(\epsilon^{\prime}\right)^{\gamma_{G^{\prime}}}\times\mbox{constant}\,.\label{eq:prop2}
\end{equation}
If (\ref{eq:prop2}) holds for any $\vec{w}^{\prime}$, we will be
able to find a constant $M^{\prime}>0$ such that 
\[
\left|F_{N}^{(g)}\left(\epsilon_{0}(\vec{w}_{0}+\epsilon^{\prime}\vec{w}^{\prime}),\vec{y}\right)\right|<\epsilon_{0}^{\gamma_{G}}\left(\epsilon^{\prime}\right)^{\gamma_{G^{\prime}}}M^{\prime}\,,
\]
which implies we can find $M>0$ satisfying the inequality (\ref{eq:inequal})
also in the neighborhood of $\vec{w}=\vec{w}_{0}$. Therefore we need
to study the case in which $(\epsilon_{0}(\vec{w}_{0}+\epsilon^{\prime}\vec{w}^{\prime}),\vec{y})$
corresponds to a singular configuration, in order to prove the continuity
of $F_{N}^{(g)}(\vec{x})$. The behavior of $F_{N}^{(g)}\left(\epsilon_{0}(\vec{w}_{0}+\epsilon^{\prime}\vec{w}^{\prime}),\vec{y}\right)$
at a possible singularity $\vec{w}^{\prime}=\vec{w}_{1}$ corresponding
to $G^{\prime\prime}\subset G^{\prime}$ can be studied by estimating
\[
F_{N}^{(g)}\left(\epsilon_{0}\vec{w}_{0}+\epsilon_{0}\epsilon_{1}\vec{w}_{1}+\epsilon_{0}\epsilon_{1}\epsilon^{\prime\prime}\vec{w}^{\prime\prime},\vec{y}\right)\,,
\]
for $\epsilon^{\prime\prime}\ll1$ with $\epsilon_{0},\epsilon_{1},\vec{w}^{\prime\prime}$
defined in the same way. 

After repeating this procedure a finite number of times, we end up
with 
\[
F_{N}^{(g)}\left(\epsilon_{0}\vec{w}_{0}+\epsilon_{0}\epsilon_{1}\vec{w}_{1}+\cdots+\epsilon_{0}\epsilon_{1}\cdots\epsilon^{(n)}\vec{w}^{(n)},\vec{y}\right)\sim\epsilon_{0}^{\gamma_{G}}\left(\epsilon_{1}\right)^{\gamma_{G^{\prime}}}\cdots\left(\epsilon^{(n)}\right)^{\gamma_{G^{(n)}}}\times\mbox{constant}
\]
 which holds for any $\vec{w}^{(n)}$, because $G$ involves only
a finite number of collapsing necks and interaction points. Applying
this procedure to all possible singularities in the neighborhood of
$(\vec{0},\vec{y})$, we can show that (\ref{eq:inequal}) holds for
any $\vec{w}$ and therefore $F_{N}^{(g)}(\vec{x})$ is continuous
at $\vec{x}=(\vec{0},\vec{y})$. 

Thus we have shown that $F_{N}^{(g)}(\vec{x})$ is continuous at possible
singularities. Since $F_{N}^{(g)}(\vec{x})$ is a differentiable function
of $\vec{x}$ away from these points, $F_{N}^{(g)}(\vec{x})$ is a
continuous function of $\vec{x}$ without any singularities. Therefore
the amplitude $A_{N}^{(g)}$ becomes finite if we choose $Q^{2}>10$
and $\varepsilon>0$, because the parameters $T_{I}$ are cut off
by the $i\varepsilon$ prescription.

\section{Discussions\label{sec:Discussions}}

In this paper, we have studied the divergences we encounter in perturbative
expansion of the amplitudes in the light-cone gauge superstring field
theory. From the point of view of light-cone gauge string field theory,
they originate from both infrared and ultraviolet regions with respect
to the world sheet metric $ds^{2}=d\rho d\bar{\rho}$ and collisions
of interaction points. The contributions from the infrared region
can be dealt with by introducing the Feynman $i\varepsilon$. In order
to regularize other kinds of divergences, we formulate the theory
in linear dilaton background. We have shown that the light-cone gauge
superstring field theory with $Q^{2}>10$ and $\varepsilon>0$ is
free from divergences at least perturbatively. 

The theory with $Q\ne0$ with eight transverse directions is not a
theory in the critical dimension and the Lorentz invariance should
be broken. However it corresponds to a conformal gauge world sheet
theory with nonstandard longitudinal part \cite{Baba:2009ns,Baba:2009fi,Ishibashi:2013nma,Ishibashi2016}
which obviously breaks the Lorentz invariance. Including the ghosts,
the total central charge of the world sheet theory is $0$ and it
is possible to construct nilpotent BRST charge. Therefore the gauge
invariance of superstring theory is not broken by making $Q\ne0$
for regularization.

It should be possible to obtain the amplitudes in the critical dimension
by defining them as analytic functions of $Q$ in the region $Q^{2}>10$
and analytically continuing them to $Q=0$ as is usually done in dimensional
regularization. In a recent paper \cite{Ishibashi2016a}, we have compared the
results with those \cite{Sen2015,Saroja1992,Sen2015a} obtained by
the first quantized formalism and shown that they coincide exactly, 
in the case of the amplitudes for even spin structure with external lines in the (NS,NS) sector. %

In this paper, we have dealt with superstring theory in Minkowski
spacetime. The results in this paper hold also for the case where
besides $X^{1}$ for the linear dilaton background the world sheet
theory consists of nontrivial conformal field theories, provided possible
singularities of $F_{N}^{(g)}$ arise only from degenerations and
collisions of interaction points. We expect that this is the case
for reasonable unitary world sheet theory. It is known that $F_{N}^{(g)}$
can become singular%
\footnote{It is possible \cite{Sen2015b,Sen2015l,Sen2016} to deal with this
kind of singularities in string field theory by introducing stubs
\cite{Zwiebach1993a}.%
} at some regular point in the interior of the moduli space because
the correlation functions involves theta functions in the denominator
\cite{Verlinde1987b}, if the world sheet theory involves nonunitary
conformal field theory like superconformal ghost. Therefore it seems
difficult to formulate a Lorentz covariant generalization of the results
in this paper. 

The regularization discussed in this paper looks similar to the dimensional
regularization in field theory but there are several crucial differences.
Firstly, the number of transverse directions, and accordingly those
of spacetime momenta and gamma matrices are fixed in our formulation.
The divergences are regularized not by reducing the number of integration
variables. We do not encounter problems with spacetime fermions, like
those in the dimensional regularization in field theory. There are
no difficulties in dealing with the amplitudes corresponding to world
sheets with odd spin structure. Since the world sheet theory involves
sixteen or eight fermionic variables, it is possible to recast the
string field theory into that in the Green-Schwarz formalism \cite{Mandelstam:1985wh}.
Secondly, we have a concrete Hamiltonian or action describing the
theory with $Q$, contrary to the case of dimensional regularization
in which there does not exist any concrete theory in fractional dimensions.
Hence the regularization proposed in this paper shall be useful in
discussing nonperturbative questions in superstring theory, although
the Hamiltonian is complex because of the dilaton background.

\section*{Acknowledgments}

We would like to thank K. Murakami for discussion. This work was supported
in part by Grant-in-Aid for Scientific Research (C) (25400242) from
MEXT.

\bibliographystyle{utphys}
\bibliography{SFTOct05_14}

\end{document}